\documentclass[prd,twocolumn,showpacs,superscriptaddress,nofootinbib,floatfix,showkeys,10pt]{revtex4-1}
\usepackage{graphicx}
\usepackage{amsmath}
\usepackage{bm}
\usepackage{yhmath}
\usepackage{mathtools}
\usepackage{wasysym}
\usepackage[colorlinks,citecolor=blue,urlcolor=blue,linkcolor=blue]{hyperref}
\usepackage{subfigure}
\usepackage{color}
\usepackage{cases}
\usepackage{subfigure}
\usepackage{times}
\usepackage{dcolumn,booktabs,bm}
\usepackage{slashed}
\usepackage{amsfonts,amssymb,stmaryrd,latexsym,amsmath}
\usepackage{textcomp}
\usepackage{multirow}
\usepackage{cancel}
\usepackage{array}

\def\SU3{{\text{SU(3)}_F}}

\renewcommand{\arraystretch}{1.8}
\begin{document}

\title{Predicting the $\bar{D}_s^{(*)} D_s^{(*)}$ bound states as the partners of $X(3872)$}

\author{Lu Meng}
\affiliation{Ruhr-Universit\"at Bochum, Fakult\"at f\"ur Physik und
Astronomie, Institut f\"ur Theoretische Physik II, D-44780 Bochum,
Germany }

\author{Bo Wang}\email{bo-wang@pku.edu.cn}
\affiliation{Center of High Energy Physics, Peking University,
Beijing 100871, China} \affiliation{School of Physics and State Key
Laboratory of Nuclear
    Physics and Technology, Peking University, Beijing 100871,
    China}

\author{Shi-Lin Zhu}\email{zhusl@pku.edu.cn}
\affiliation{School of Physics and State Key Laboratory of Nuclear
Physics and Technology, Peking University, Beijing 100871,
China}\affiliation{Center of High Energy Physics, Peking University,
Beijing 100871, China}

\begin{abstract}
In this work, we investigate the SU(3) flavor symmetry, heavy quark
spin symmetry and their breaking effects in the di-meson systems. We
prove the existence of the $[\bar{D}_{s}^{*}D_{s}^{*}]^{0^{++}}$,
$[\bar{D}_{s}^{*}D_{s}/\bar{D}_{s}^{}D_{s}^*]^{1^{+-}}$, and
$[\bar{D}_{s}^{*}D_{s}^{*}]^{1^{+-}}$ bound states as the
consequence of two prerequisites in the SU(3) flavor symmetry and
heavy quark spin symmetry. The first prerequisite, the $X(3872)$ as
the weakly $\bar{D}^{*}D/\bar{D} D^{*}$ bound state is supported by
its mass and decay branching ratios. The second prerequisite, the
existence of the $[\bar{D}_{s}D_{s}]^{0^{++}}$ bound state is
supported by the lattice QCD calculation~\cite{Prelovsek:2020eiw}
and the observation of $\chi_{c0}(3930)$ by the LHCb
Collaboration~\cite{Aaij:2020hon,Aaij:2020ypa}. We hope the future
experimental analyses can search for these bound states in the $B\to
D_{(s)}^{(*)}\bar{D}^{(*)}_{(s)}h$ processes ($h$ denotes the light
hadrons). The $[\bar{D}_{s}^{*}D_{s}^{*}]^{0^{++}}$ bound state is
also expected to be reconstructed in the $J/\psi \phi$ final state
in the $B\to J/\psi \phi K$ decay.
\end{abstract}

\keywords{SU(3) flavor symmetry; hadronic molecule; $X(3872)$}

\maketitle

\thispagestyle{empty}

\section{Introduction}\label{sec:intro}
The SU(3) flavor [$\SU3$] symmetry is an approximate symmetry of QCD
Lagrangian, which manifests itself in the hadron spectra. From the
observation of $\Omega^-$ in the 1960s~\cite{Barnes:1964pd}, the
SU(3)$_F$ symmetry was well used to classify the mesons and baryons
into multiplets. However, for a long time, the SU(3)$_F$ symmetry
was seldom investigated in the superstructures of QCD, such as the
di-meson systems.

Since the observation of $X(3872)$ in 2003~\cite{Choi:2003ue}, more
and more hidden charm/bottom exotic hadrons have been reported in
experiments~\cite{Brambilla:2019esw,Liu:2019zoy,Guo:2017jvc,Olsen:2017bmm,Chen:2016qju,Esposito:2016noz}.
Many of these states are in the proximity of di-hadron thresholds.
For example, the $X(3872)$~\cite{Choi:2003ue},
$Z_c(3900)$~\cite{Belle:2011aa},
$Z_c(4020)$~\cite{Ablikim:2013wzq,Ablikim:2013emm} and
$P_c$~\cite{Aaij:2015tga,Aaij:2019vzc} states are near the
$\bar{D}^{*0}D^0/\bar{D}^0 D^{*0}$, $\bar{D}^*D/\bar{D}D^*$,
$\bar{D}^*D^*$ and $\Sigma_c\bar{D}^{(*)}$ thresholds, respectively,
which indicates these states might be the di-hadron bound states or
resonances. Recently, the strange hidden charm pentaquark candidate
$P_{cs}(4459)^0$ was reported by the LHCb
Collaboration~\cite{Aaij:2020gdg}, which is in good agreement with
our previous prediction of the $\Xi_c\bar{D}^*$ bound state with a
mass $4456.9$ MeV~\cite{Wang:2019nvm}. Very recently, the
$Z_{cs}(3985)^-$ state was observed by the BESIII
Collaboration~\cite{Ablikim:2020hsk}, which could be interpreted as
the $\bar{D}_s D^*/\bar{D}^*_{s}D$ di-meson states as the $U$-spin
partner of $Z_c(3900)^-$~\cite{Meng:2020ihj,Wang:2020htx}. The
$P_{cs}(4459)^0$ and $Z_{cs}(3985)^-$ states are good candidates of
the strange partners of $P_c$ and $Z_c(3900)$ states in the $\SU3$
symmetry, which inspired many works about the $\SU3$ symmetry for
the di-hadron
systems~\cite{Wang:2020rcx,Sun:2020hjw,Du:2020vwb,Cao:2020cfx,Chen:2020yvq,Yang:2020nrt,Wang:2020kej,Azizi:2020zyq,Jin:2020yjn,Wan:2020oxt,Liu:2020hcv,Chen:2020kco,Peng:2020hql,Chen:2020uif,Shen:2020gpw,Stancu:2020paw,Xu:2020evn,Ikeno:2020csu}.

The heated discussion on the strange partners of the $P_c$ and $Z_c$
states reminds us of the $\SU3$ partner states of $X(3872)$, the
super star in the exotic hadron family. In
Refs.~\cite{Nieves:2012tt,Guo:2013sya,Baru:2016iwj}, the heavy quark
spin symmetry (HQSS) partners of $X(3872)$ with the $J^{PC}$ quantum
numbers $2^{++}$ was proposed. In this work, we will consider the
$\bar{D}_s^{(*)} D_s^{(*)}$ states as the counterparts of $X(3872)$
in the HQSS and $\SU3$ symmetry. Recently, the lattice QCD
calculation with $m_\pi \simeq$ 280 MeV indicated the existence of
the scalar $\bar{D}_s D_s$ bound state~\cite{Prelovsek:2020eiw},
which might correspond to the $\chi_{c0}(3930)$ observed by the LHCb
Collaboration~\cite{Aaij:2020ypa,Aaij:2020hon}. In this work, we
will show that the existence of the $\bar{D}_s^{(*)} D_s^{(*)}$
bound states is the natural consequence of two prerequisites in the
$\SU3$ symmetry and HQSS,
\begin{itemize}
    \item The $X(3872)$ is the molecular state with its mass coinciding exactly with the $\bar{D}^*_0{D}_0$
    threshold;
    \item There exist the $\bar{D}_sD_s$ bound states with $J^{PC}=0^{++}$.
\end{itemize}

This work is organized as follows. In Sec.~\ref{sec:symmtry}, we
discuss the $\SU3$ symmetry, HQSS and their breaking effects for the
$\bar{D}_{(s)}^{(*)} D_{(s)}^{(*)}$ di-meson systems. In
Sec.~\ref{sec:dsds}, we prove the existence of $\bar{D}_s^{(*)}
D_s^{(*)}$ bound states from the perspective of $\SU3$ symmetry and
HQSS as the partners of $X(3872)$. In Sec.~\ref{sec:cc}, we show
that the predictions of the $\bar{D}_{(s)}^{(*)} D_{(s)}^{(*)}$
bound states are valid when the coupled-channel formalism is adopted
for $X(3872)$. We conclude with a brief discussion and a short
summary in Sec.~\ref{sec:sum}.

\section{$\SU3$ symmetry and HQSS for di-meson systems}\label{sec:symmtry}

\begin{figure}[!htp]
    \centering
    \includegraphics[width=0.45\textwidth]{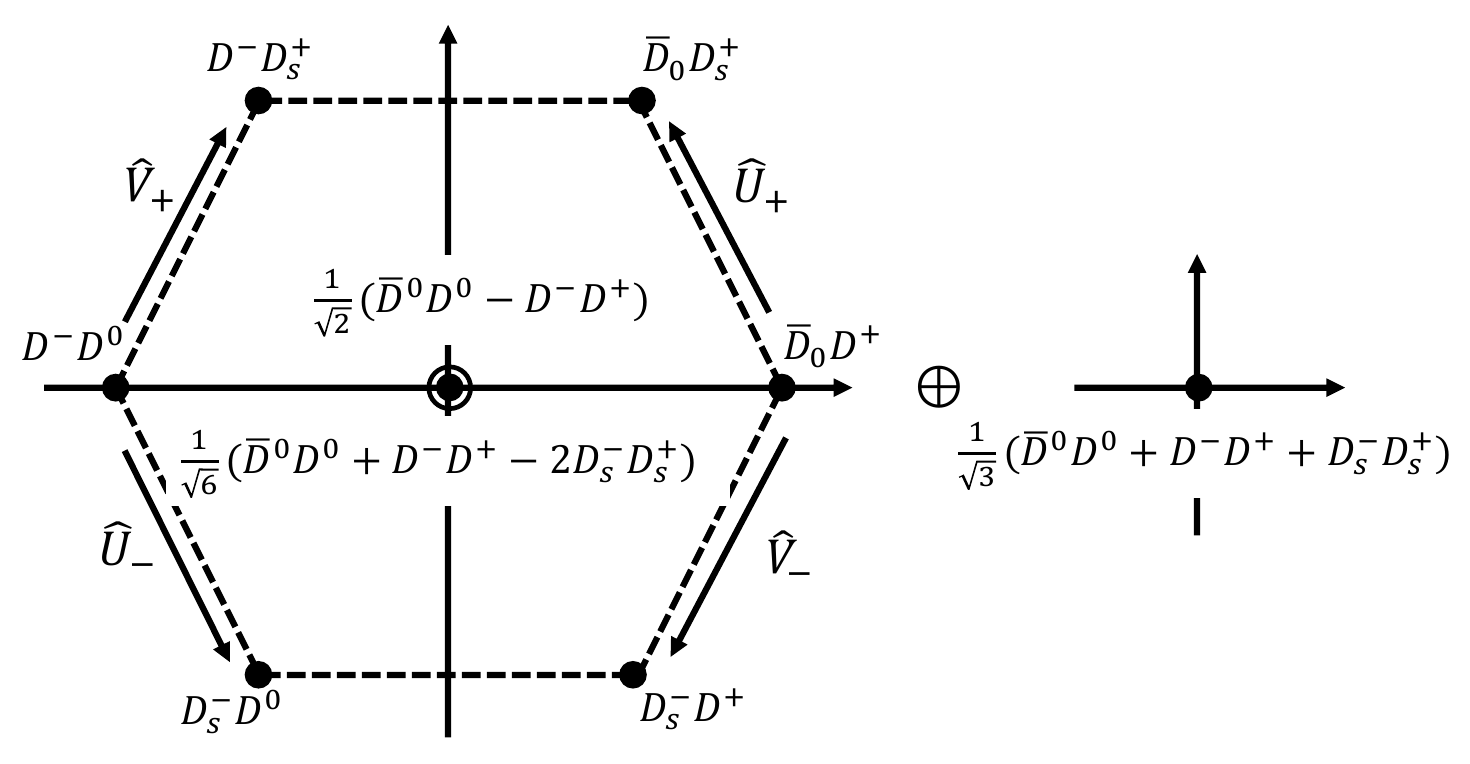}
    \caption{The multiplet structure of $\bar{D}_{(s)}{D}_{(s)}$ di-meson systems in the $\SU3$ symmetry, where the $\hat{U}_\pm$ and $\hat{V}_\pm$ denote the $U$-spin and $V$-spin ladder operators~\cite{Meng:2020ihj}.}\label{fig:su3}
\end{figure}

We will take the $\bar{D}_{(s)}^{(*)}D_{(s)}^{(*)}$ as an example to
discuss the SU(3)$_F$ symmetry and HQSS for the di-meson systems. In
Fig.~\ref{fig:su3}, we present the $\bar{D}_{(s)}D_{(s)}$ multiplets
and their flavor wave functions in the $\SU3$ symmetry. Their other
HQSS partners have the similar structures. We can see that the
hidden strange $\bar{D}_s D_s$ system will mix with the $\bar{D}D$
system in the $\SU3$ limit. For the spin wave function, we list the
inner products of HQSS basis and di-meson basis in
Table~\ref{tab:hq}. In the heavy quark limit, the mixture also
occurs in the two $0^{++}$ di-meson states and two $1^{+-}$ di-meson
states, respectively. The $C$-parity is only for the neutral states
here and below.

For the charmed mesons, the breaking effects will result in two
kinds of mass splittings,
 \begin{eqnarray}
    D_{s}^{(*)}-D^{(*)}&\simeq100\text{ MeV},\label{eq:mssu3}
    \\D_{(s)}^{*}-D_{(s)}&\simeq140\text{ MeV},\label{eq:masshq}
 \end{eqnarray}
which arises from the $\SU3$ symmetry and HQSS breaking effects,
respectively. However, the observed di-meson candidates are in the
proximity of thresholds about several MeVs, which are much smaller
than the mass splittings in Eqs.~\eqref{eq:mssu3} and
\eqref{eq:masshq}. In other words, the interactions accounting for
these molecular di-meson systems are too weak to lead to significant
mixture between states with mass difference over 100 MeV. Thus, when
the large mass splittings in Eqs.~\eqref{eq:mssu3} and
\eqref{eq:masshq} are involved, the states in the real world will be
distinguishable according to the di-meson thresholds rather than the
$\SU3$ symmetry and HQSS.

\begin{table*}
\centering
\renewcommand{\arraystretch}{1.5}
    \caption{The HQSS basis, di-meson basis and their inner product in the spin space. In the heavy quark symmetry basis, the spin wave functions are denoted as $|S_{H}^{P_H C_H},S_{L}^{P_L C_L};J^{PC}\rangle$, where the subscript $H/L$ represents the heavy/light degree of freedom ($C$-parity only for the neutral states).}\label{tab:hq}
    \setlength{\tabcolsep}{1.3mm}
    \begin{tabular}{ccccccc}
        \hline \hline
        & $|\bar{D}D;0^{++}\rangle$ & $|\frac{1}{\sqrt{2}}(\bar{D}^{*}D+\bar{D}D^{*});1^{+-}\rangle$ & $|\frac{1}{\sqrt{2}}(\bar{D}^{*}D-\bar{D}D^{*});1^{++}\rangle$ & $|\bar{D}^{*}D^{*};0^{++}\rangle$ & $|\bar{D}^{*}D^{*};1^{+-}\rangle$ & $|\bar{D}^{*}D^{*};2^{++}\rangle$\tabularnewline
        \hline
        $|0_{H}^{-+},0_{L}^{-+};0^{++}\rangle$ & $-\frac{1}{2}$ &  &  & $\frac{\sqrt{3}}{2}$ &  & \tabularnewline
        $|0_{H}^{-+},1_{L}^{--};1^{+-}\rangle$ &  & $\frac{1}{\sqrt{2}}$ &  &  & $\frac{1}{\sqrt{2}}$ & \tabularnewline
        \hline
        $|1_{H}^{--},0_{L}^{-+};1^{+-}\rangle$ &  & $-\frac{1}{\sqrt{2}}$ &  &  & $\frac{1}{\sqrt{2}}$ & \tabularnewline
        $|1_{H}^{--},1_{L}^{--};0^{++}\rangle$ & $-\frac{\sqrt{3}}{2}$ &  &  & $-\frac{1}{2}$ &  & \tabularnewline
        $|1_{H}^{--},1_{L}^{--};1^{++}\rangle$ &  &  & $1$ &  &  & \tabularnewline
        $|1_{H}^{--},1_{L}^{--};2^{++}\rangle$ &  &  &  &  &  & 1\tabularnewline
        \hline \hline
    \end{tabular}
\end{table*}

Unlike the di-meson thresholds, it is reasonable to presume the
interactions between two mesons satisfy the $\SU3$ and HQSS. We
define the relative energy $\Delta E$ and binding energy $E_b$ of
di-meson system as
$$\Delta E=-E_b= M_{\text{dimeson}}-M_{\text{threshold}}.$$
The $\Delta E$ of the di-meson systems are at the order of 10 MeV.
The relative momentum is at the order of $\sqrt{|\Delta E|
m_{D_s^{(*)}}}\sim 130$ MeV, which is a small scale compared to the
charm quark mass. Thus, the interactions between two mesons are very
soft, which manifests the heavy quark symmetry. The interactions
between mesons originating from either the flavor-blind gluonic
interaction or exchanging $\SU3$ multiplet mesons will result in the
approximate $\SU3$ symmetry. Thus, in this work, we will consider
the $\SU3$ symmetry and HQSS breaking effects in constructing the
spin and flavor functions but take the symmetry limits in modeling
the hadronic interaction.

We could embed the HQSS and $\SU3$ symmetry with the interactions at
the quark level,
\begin{equation}
    V_{q\bar{q}}=c_{1}+c_{2}\bm{s}_{1}\cdot\bm{s}_{2}+c_{3}\mathbb{C}_{2}+c_{4}(\bm{s}_{1}\cdot\bm{s}_{2})\mathbb{C}_{2},\label{eq:vqq}
\end{equation}
where $V_{q\bar{q}}$ denotes the interactions between the light
quark and antiquark [The interactions involving the heavy
(anti)quark is suppressed in the heavy quark symmetry, which is
neglected]. The $ \bm{s}_i$ is the spin operator of the light
(anti)quark. $\mathbb{C}_2=-\sum_{i=1}^8 \lambda_F^i\lambda_F^{*i}$
is the Casimir operator in the flavor space. Eq.~\eqref{eq:vqq}
represents the general interaction of the $S$-wave
$\bar{D}_{(s)}^{(*)}{D}_{(s)}^{(*)}$ channel in the $\SU3$ symmetry
and HQSS, which could be realized in other equivalent
approaches~\cite{Meng:2019ilv,Wang:2019ato}.

\section{$X(3872)$ as the $\bar{D}^{*0}D^0/\bar{D}^0 D^{*0}$ molecular state}\label{sec:dsds}

The $X(3872)$ mass coincides exactly with the $\bar{D}^{*0}{D}^0$
threshold~\cite{Zyla:2020zbs},
\begin{equation}
    m_{D^0}+m_{D^{*0}}-m_{X(3872)}=(0.00\pm0.18) \text{ MeV}.
\end{equation}
Meanwhile, it has the large branching faction of $X(3872)\to
\bar{D}^{*0}D^0$~\cite{Zyla:2020zbs}. Its mass and branch fraction
indicate that the main component of $X(3872)$ is
$\bar{D}^{*0}D^0/\bar{D}^0 D^{*0}$. The scarcity of
${D}^{*-}D^+/\bar{D}^- D^{*+}$ component induce to large isospin
violation effect. The large ratio of the branching fraction
$\mathcal{B}(X\to J/\psi \omega)/\mathcal{B}( X\to J/\psi
\pi^+\pi^-)$ also support that the $X(3872)$ is not an eigenstate of
isospin~\cite{delAmoSanchez:2010jr}. The $D^{*-}D^+/D^{-}D^{*+}$
threshold is above the $X(3872)$ by about $8$ MeV, which could be
regarded as a large scale compared with the width and binding energy
of $X(3872)$ as the $\bar{D}^{*0}D^0/\bar{D}^0 D^{*0}$ bound state.
Thus, it is very natural to interpret the $X(3872)$ as the
$\bar{D}^{*0}D^0/\bar{D}^0 D^{*0}$ di-meson system. Similar
perspectives were addressed in the refined one-boson-exchanged
calculation~\cite{Li:2012cs} and XEFT~\cite{Fleming:2007rp}. In this
section, we will treat the $X(3872)$ as the
$\bar{D}^{*0}D^0/\bar{D}^0 D^{*0}$ molecular state.

As a $\bar{D}^{*0}D^0/\bar{D}^0 D^{*0}$ di-meson system, the flavor
wave function of $X(3872)$ in the light part will be
$|u\bar{u}\rangle$. As a consequence, it has the same flavor matrix
element with the $\bar{D}^{(*)}_sD^{(*)}_s$ systems,
\begin{equation}
    \langle\mathbb{C}_{2}\rangle_{u\bar{u}}=\langle\mathbb{C}_{2}\rangle_{s\bar{s}}.\label{eq:equuss}
\end{equation}
In other words, the $\bar{D}^{(*)}_sD^{(*)}_s$ systems are the
$V$-spin partners of the $X(3872)$~\cite{Meng:2020ihj}. Thus, we
could reparameterize the interactions of $X(3872)$ and
$\bar{D}^{(*)}_sD^{(*)}_s$ systems as,
\begin{equation}
    V_{q\bar{q}}=\tilde{c}_1+\tilde{c}_2\bm{s}_1\cdot\bm{s}_2,\label{eq:vqq2}
\end{equation}
where their common flavor information has been absorbed into the new
coupling constants $\tilde{c}_1$ and $\tilde{c}_2$.

In the spin space, the matrix elements read~\cite{Meng:2020ihj},
\begin{eqnarray}
    &&\langle\bm{s}_{1}\cdot\bm{s}_{2}\rangle_{\{\mathtt{PP},\mathtt{VV}\}}^{0^{++}}=\left[\begin{array}{cc}
        0 & \frac{\sqrt{3}}{4}\\
        \frac{\sqrt{3}}{4} & -\frac{1}{2}
    \end{array}\right],\label{eq:ll0}\\
    &&\langle\bm{s}_{1}\cdot\bm{s}_{2}\rangle_{\{\mathtt{PV},\mathtt{VV}\}}^{1^{+-}}=\left[\begin{array}{cc}
        -\frac{1}{4} & -\frac{1}{2}\\
        -\frac{1}{2} & -\frac{1}{4}
    \end{array}\right],\label{eq:ll}\\
    &&\langle\bm{s}_{1}\cdot\bm{s}_{2}\rangle_{\{\mathtt{PV}\}}^{1^{++}}=\frac{1}{4},\quad
    \langle\bm{s}_{1}\cdot\bm{s}_{2}\rangle_{\{\mathtt{VV}\}}^{2^{++}}=\frac{1}{4},\label{eq:ll2}
\end{eqnarray}
where $\mathtt{P}$ and $\mathtt{V}$ represent the pseudoscalar and
vector heavy mesons, receptively. The superscript denotes the
$J^{PC}$ of the di-meson channel. Though we present the off-diagonal
matrix elements, their mixing effect is negligible as discussed in
Sec.~\ref{sec:symmtry}. We can easily obtain the following relation,
\begin{eqnarray}
&   \left(V_{\mathtt{PV}}^{1^{++}}-V_{\mathtt{PP}}^{0^{++}}\right):\left(V_{\mathtt{VV}}^{0^{++}}-V_{\mathtt{PP}}^{0^{++}}\right):\left(V_{\mathtt{PV/VV}}^{1^{+-}}-V_{\mathtt{PP}}^{0^{++}}\right)\nonumber \\
&~~~~~~~~~~~~~~~~~~~~~~~~~~~~~~~~~~~~~~~~~~~~~
=1:-2:-1.\label{eq:ratio}
\end{eqnarray}

\begin{figure}
 \centering
\includegraphics[width=0.48\textwidth]{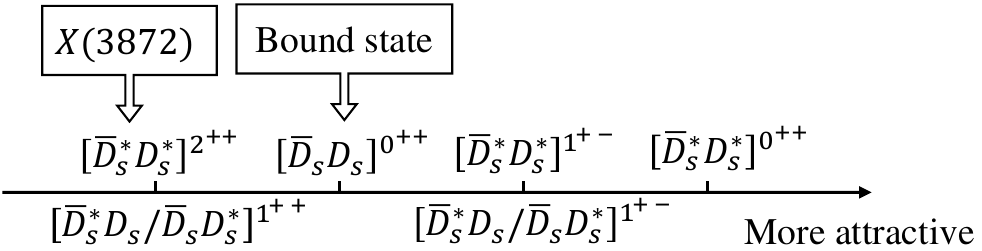}
\caption{The interactions of $\bar{D}_s^{(*)}D_s^{(*)}$ systems in
the order of becoming more attractive.}\label{fig:inter_order}
\end{figure}

The recent lattice QCD calculation yielded a shallow
$[\bar{D}_sD_s]^{0^{++}}$ bound state with $\Delta
E=-6.2^{+3.8}_{-2.0}$\text{ MeV}~\cite{Prelovsek:2020eiw}. The
$X(3872)$ is a marginal bound state and its binding energy is less
than $1$ MeV. Since the $\bar{D}_sD_s$ is a deeper bound state than
$X(3872)$, the interaction in the $[\bar{D}_sD_s]^{0^{++}}$ channel
is more attractive than that in the $[\bar{D}^{*0}D^0/\bar{D}^0
D^{*0}]^{1^{++}}$ channel. According to
Eqs.~\eqref{eq:ll0}-\eqref{eq:ll2}, the $[\bar{D}_sD_s]^{0^{++}}$
channel is more attractive than
$[\bar{D}^*_sD_s/D^*_s\bar{D}_s]^{1^{++}}$ one, which indicates the
coupling constant in Eq.~\eqref{eq:vqq2} $\tilde{c}_2 >0$. Thus, we
obtain the order of interactions for $D_s^{(*)}\bar{D}_s^{(*)}$ in
Fig.~\ref{fig:inter_order}, which become more attractive along the
arrow. As a consequence, the more attractive interactions in the
$[\bar{D}_s^* D_s/\bar{D}_s D^*_s]^{1^{+-}}$,
$[\bar{D}_s^{*}D_s^*]^{1^{+-}}$ and $[\bar{D}_s^{*}D_s^*]^{0^{++}}$
will give rise to deeper binding solutions. Among them, the
$[\bar{D}_s^{*}D_s^*]^{0^{++}}$ bound state is the deepest one.

Apart from the evidence of the $\bar{D}_s D_s$ bound states in
lattice QCD, the $\chi_{c0}(3930)$ observed by LHCb collaboration
was a good candidate in experiment~\cite{Aaij:2020hon,Aaij:2020ypa}.
The $\chi_{c0}(3930)$ is below the $\bar{D}_s D_s$ threshold about
$12.9$ MeV, which will lead to the deeper $[\bar{D}_s^*
D_s/\bar{D}_s D^*_s]^{1^{+-}}$, $[\bar{D}_s^{*}D_s^*]^{1^{+-}}$ and
$[\bar{D}_s^{*}D_s^*]^{0^{++}}$ binding solutions. The existence of
the $[\bar{D}_s^* D_s/\bar{D}_s D^*_s]^{1^{+-}}$,
$[\bar{D}_s^{*}D_s^*]^{1^{+-}}$ and $[\bar{D}_s^{*}D_s^*]^{0^{++}}$
bound states arises from the $\SU3$ symmetry and HQSS, which is
independent of the specific interaction mechanism.

%\section{Numerical calculation}\label{sec:cal}
We adopt a simple model to demonstrate the argument proposed in
Sec.~\ref{sec:dsds} numerically. We assume the hadronic interaction
of the $X(3872)$ and $\bar{D}_s^{(*)} D_s^{(*)}$ systems are contact
interaction, as shown in Eq.~\eqref{eq:vqq2}. We introduce the
$g(\bm{p})=\exp ({-\bm{p}^2/\Lambda^{2}})$ to regulate divergence in
Lippmann-Schwinger equations (LSEs),
\begin{equation}
    T(\bm{p}',\bm{p};E)=V(\bm{p}',\bm{p})+\int\frac{d^{3}\bm{p}''}{(2\pi)^{3}}\frac{V(\bm{p}',\bm{p}'')T(\bm{p}'',\bm{p};E)}{E-\bm{p^2}/(2\mu)+i\epsilon},~\label{eq:ls}
\end{equation}
where $\mu $ is the reduced mass. The potential reads
\begin{equation}
V(\bm{p}',\bm{p}'')=v g(\bm{p}')g(\bm{p}''),\label{eq:sep_v}
\end{equation}
where $v$ is the coupling constant. In the calculation, we take the
physical masses of the corresponding channels~\cite{Zyla:2020zbs}.
With substitution $ T(\bm{p}',\bm{p};E)=t(E)g(\bm{p}')g(\bm{p}'')$
in Eq.~\eqref{eq:ls}, we obtain the algebraic equation,
\begin{equation}
    t(E)=v+vF(E)t(E),\label{eq:ls_1cn}
\end{equation}
where
\begin{equation}
    F(E)=\int_{0}^{\infty}dp''\frac{4\pi}{(2\pi)^{3}}\frac{2\mu p''^{2}g(\bm{p}'')^{2}}{2\mu E-p''^{2}+i\epsilon}.
\end{equation}
  The solution is
  \begin{equation}
    t^{-1}(E)=a(\Lambda)-F(E;\Lambda),\text{ with } a(\Lambda)\equiv {1\over v(\Lambda)},
  \end{equation}
where the cutoff-dependence of $a$ and $F$ cancels out. We can
obtain the pole of $t(E)$ below the threshold which corresponds to
the bound state.

In this work, we choose the cutoff $\Lambda =1.0$ GeV. One can
choose different regulators and cutoffs, or alternatively solve the
Schr\"odinger equation~\cite{Meng:2019ilv,Wang:2019ato} in
coordinate space, which gives the same physical implications. We
choose the $0$ and $1$ MeV as the lower and upper limits of the
binding energy of $X(3872)$, and adopt the binding energy of the
$[\bar{D}_{s}D_{s}]^{0^{++}}$ in the range of lattice
QCD~\cite{Prelovsek:2020eiw} and experimental
measurement~\cite{Aaij:2020hon,Aaij:2020ypa} as inputs. We present
the predictions of the $[\bar{D}_{s}^{*}D_{s}^{*}]^{0^{++}}$,
$[\bar{D}_{s}^{*}D_{s}/\bar{D}_{s}^{}D_{s}^*]^{1^{+-}}$, and
$[\bar{D}_{s}^{*}D_{s}^{*}]^{1^{+-}}$ systems in Table~\ref{tab:be}.
There do exist the bound states for these systems with the binding
energy from several MeVs to several tens of MeVs. One can see that
they are bound more deeply than the $[\bar{D}_{s}D_{s}]^{0^{++}}$
state. Among them, the $[\bar{D}_{s}^{*}D_{s}^{*}]^{0^{++}}$ is the
deepest one. The mass ranges of the
$[\bar{D}_{s}^{*}D_{s}^{*}]^{0^{++}}$,
$[\bar{D}_{s}^{*}D_{s}/\bar{D}_{s}^{}D_{s}^*]^{1^{+-}}$ and
$[\bar{D}_{s}^{*}D_{s}^{*}]^{1^{+-}}$ systems are predicted to be
$[4140.1,4216.1]$, $[4036.8,4075.6]$ and $[4177.2,4218.1]$ MeV,
respectively. Though the binding energies are sensitive to the
input, the existence of these bound states is persistent.

\begin{figure}[!htp]
    \centering
    \includegraphics[width=0.4\textwidth]{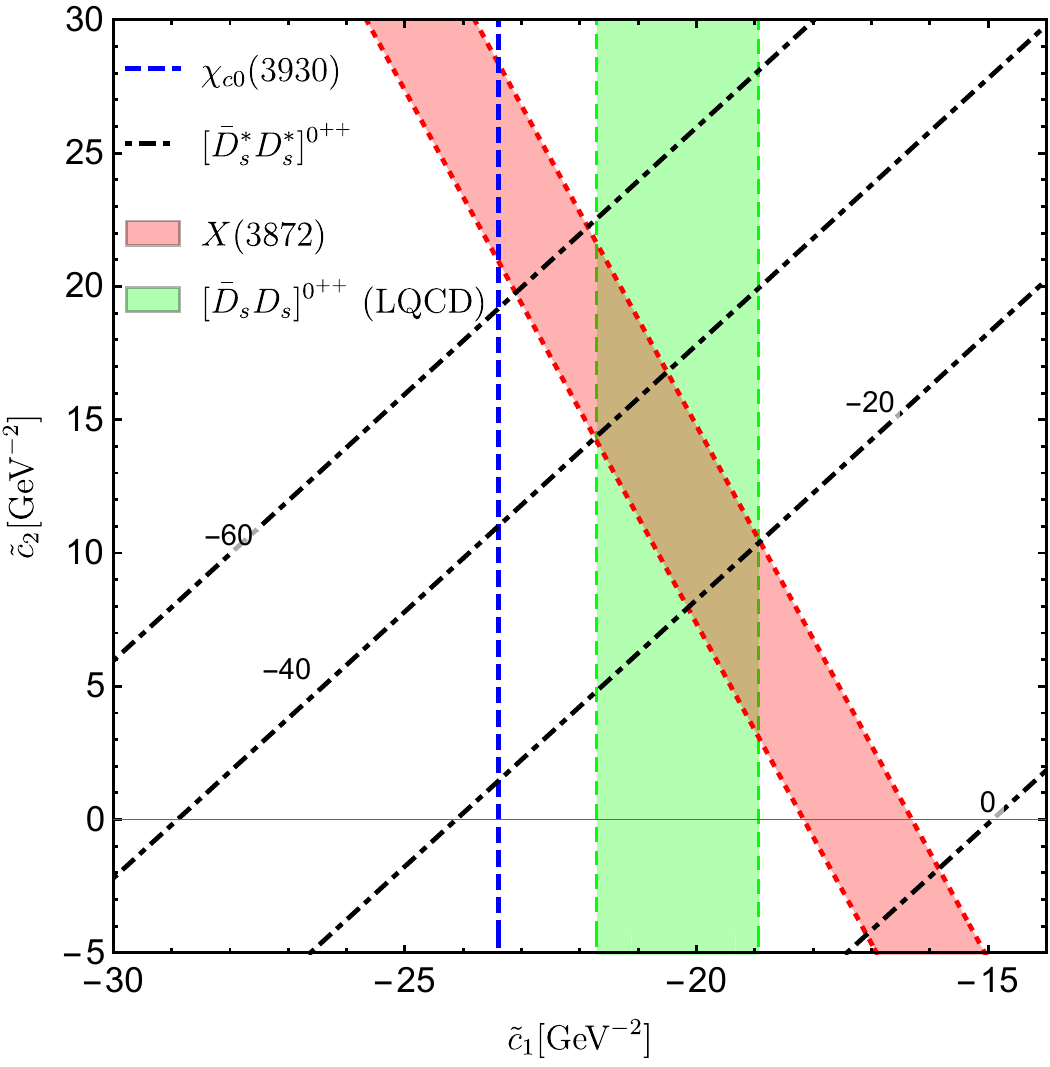}
    \caption{The parameter regions of the bound states $X(3872)$, $[\bar{D}_sD_s]^{0^{++}}$ and $[\bar{D}_s^{*}D_s^*]^{0^{++}}$. The red and green bands denote the parameter regions of $X(3872)$ with binding energy $0\sim 1$ MeV and $[\bar{D}_sD_s]^{0^{++}}$ with binding energy $2.4 \sim 8.2$ MeV~\cite{Prelovsek:2020eiw}, respectively. The dashed blue line is the parameter region of $\chi_{c0}(3930)$ as the $[\bar{D}_sD_s]^{0^{++}}$ bound states~\cite{Aaij:2020hon,Aaij:2020ypa}. The black dotdashed lines correspond to the $[\bar{D}_s^{*}D_s^*]^{0^{++}}$ with $\Delta E=0$, $-20$, $-40$ and $-60$ MeV.}\label{fig:para}
\end{figure}

We take the $[\bar{D}_{s}^{*}D_{s}^{*}]^{0^{++}}$ state as an
example to show the related parameter regions in
Fig.~\ref{fig:para}. In the overlap parameter region of the
$X(3872)$ and $[\bar{D}_{s}D_{s}]^{0^{++}}$ as the bound states, the
$[\bar{D}_{s}^{*}D_{s}^{*}]^{0^{++}}$ state is also bound. Taking
either the lattice QCD $[\bar{D}_{s}D_{s}]^{0^{++}}$
result~\cite{Prelovsek:2020eiw} or the $\chi_{c0}(3930)$ as the
$[\bar{D}_{s}D_{s}]^{0^{++}}$ bound
state~\cite{Aaij:2020hon,Aaij:2020ypa} gives the same implication.

\begin{table*}
\centering
\renewcommand{\arraystretch}{1.2}
    \caption{The $\Delta E$  and masses for the $[\bar{D}_{s}^{*}D_{s}^{*}]^{0^{++}}$, $[\bar{D}_{s}^{*}D_{s}/\bar{D}_{s}^{}D_{s}^*]^{1^{+-}}$ and $[\bar{D}_{s}^{*}D_{s}^{*}]^{1^{+-}}$ systems (the results are given in units of MeV). The binding energies of $X(3872)$ and $[\bar{D}_{s}D_{s}]^{0^{++}}$ systems are adopted as the inputs~\cite{Prelovsek:2020eiw,Aaij:2020hon,Aaij:2020ypa}. In the last two rows, we list the numerical results from Ref.~\cite{HidalgoDuque:2012pq}. }\label{tab:be}
    \setlength{\tabcolsep}{5mm}
\begin{tabular}{ccccccccc}
    \toprule[1pt]\toprule[1pt]
    $X(3872)_{\text{input}}$ & \multicolumn{2}{c}{$[\bar{D}_{s}D_{s}]_{\text{input}}^{0^{++}}$} & \multicolumn{2}{c}{$[\bar{D}_{s}^{*}D_{s}^{*}]^{0^{++}}$} & \multicolumn{2}{c}{$[\bar{D}_{s}^{*}D_{s}/\bar{D}_{s}D_{s}^{*}]^{1^{+-}}$} & \multicolumn{2}{c}{$[\bar{D}_{s}^{*}D_{s}^{*}]^{1^{+-}}$}\tabularnewline

    $\Delta E$ & $\Delta E$ & $M$ & $\Delta E$ & $M$ & $\Delta E$ & $M$ & $\Delta E$ & $M$\tabularnewline
    \midrule[1pt]
    $0.0$ & $-2.4$ & 3934.3 & $-20.3$ & $4204.1$ & $-9.5$ & $4071.0$ & $-11.4$ & $4213.0$\tabularnewline
    $0.0$ & $-6.2$ & 3930.5 & $-45.5$ & $4178.9$ & $-22.5$ & $4058.0$ & $-25.2$ & $4199.2$\tabularnewline
    $0.0$ & $-8.2$ & 3928.5 & $-57.6$ & $4166.8$ & $-29.0$ & $4051.5$ & $-32.0$ & $4192.4$\tabularnewline
    $0.0$ & -$12.9$ & 3923.8 & $-84.3$ & $4140.1$ & $-43.7$ & $4036.8$ & $-47.2$ & $4177.2$\tabularnewline
    \midrule[1pt]
    $-1.0$ & $-2.4$ & 3934.3 & $-8.3$ & $4216.1$ & $-4.9$ & $4075.6$ & $-6.3$ & $4218.1$\tabularnewline
    $-1.0$ & $-6.2$ & 3930.5 & $-28.9$ & $4195.5$ & $-15.9$ & $4064.6$ & $-18.2$ & $4206.2$\tabularnewline
    $-1.0$ & $-8.2$ & 3928.5 & $-39.6$ & $4184.8$ & $-21.7$ & $4058.8$ & $-24.4$ & $4200.0$\tabularnewline
    $-1.0$ & $-12.9$ & 3923.8 & $-64.1$ & $4160.3$ & $-35.2$ & $4045.3$ & $-38.5$ & $4185.9$\tabularnewline
    \midrule[1pt] \midrule[1pt]
    cutoff-I~\cite{HidalgoDuque:2012pq} & $-13$ & 3924 & -84 & 4140 & $-46$ & 4035 & $-47$ & 4177\tabularnewline
    cutoff-II~\cite{HidalgoDuque:2012pq} & $-9$ & 3928 & -84 & 4140 & $-41$ & 4040 & $-44$ & 4180\tabularnewline
    \bottomrule[1pt]\bottomrule[1pt]
\end{tabular}
\end{table*}

\section{$X(3872)$ in the coupled-channel formalism}\label{sec:cc}
There are different interpretations for the large isospin breaking
decays of $X(3872)$. In
Refs.~\cite{Gamermann:2009fv,Gamermann:2009uq}, the authors
interpreted the isospin breaking effect as the consequence of the
mass splitting in the propagators of the $\bar{D}^{*0}D^0/\bar{D}^0
D^{*0}$ and $D^+D^{*-}/D^-D^{*+}$ components, which is amplified by
the different effective phase spaces of $J/\psi \rho$ and $J/\psi
\omega$. In this section, we include $D^+D^{*-}/D^-D^{*+}$
components for $X(3872)$ with the coupled-channel formalism. We will
demonstrate that a less attractive (more repulsive) $[\bar{D}_s^*
D_s/\bar{D}_s D^*_s]^{1^{++}}$ interaction is obtained than in the
single-channel case. Thus, the conclusions in Sec.~\ref{sec:dsds}
are still valid.

We adopt the renormalizable effective field theory that describes
two scattering
channels~\cite{Cohen:2004kf,Braaten:2005ai,Dong:2020hxe}. We
introduce the two channel potential for $X(3872)$,
\begin{equation}
    \left[\begin{array}{cc}
        v_{11} & v_{12}\\
        v_{12} & v_{22}
    \end{array}\right]=\frac{1}{2}\left[\begin{array}{cc}
        V^{I=1}+V^{I=0} & V^{I=0}-V^{I=1}\\
        V^{I=0}-V^{I=1} & V^{I=1}+V^{I=0}
    \end{array}\right],\label{eq:cpint}
\end{equation}
where the channel 1 is the $\bar{D}^{*0}D^0/\bar{D}^0 D^{*0}$
channel and channel 2 is the $D^{*+}D^-/D^{*-}D^+$ channel.
$V^{I=0}$ and $V^{I=1}$ are the interactions for $I=0$ and $I=1$
channels, respectively. The potential in Eq.~\eqref{eq:cpint}
satisfies the isospin symmetry. We introduce the regulator
$g(\bm{p}^2)$ and adopt the separable potentials like
Eq.~\eqref{eq:sep_v}. We can reduce the coupled-channel LSEs into
algebraic equations,
\begin{equation}
    t_{ij}(E)=v_{ij}+\sum_{a=1,2}v_{ia}F_{a}t_{aj}(E),\label{eq:cplse}
\end{equation}
where
\begin{eqnarray}
    F_{1}(E)&=&\int_{0}^{\infty}dp''\frac{4\pi}{(2\pi)^{3}}\frac{2\mu p''^{2}g(\bm{p}'')^{2}}{2\mu E-p''^{2}+i\epsilon},\nonumber
    \\F_{2}(E)&=&\int_{0}^{\infty}dp''\frac{4\pi}{(2\pi)^{3}}\frac{2\mu p''^{2}g(\bm{p}'')^{2}}{2\mu(E-\delta)-p''^{2}+i\epsilon}.\label{eq:cpf}
\end{eqnarray}
 $\delta =m_{D^{*+}}+m_{D^-}-m_{D^{*0}}-m_{\bar{D}^0}\approx 8$ MeV. For convenience, we introduce the hard regulator $g(\bm{p})=\theta(\Lambda-p)$. For $E<0$, we obtain
\begin{equation}
    F_{1}=\frac{8\mu\pi}{(2\pi)^{3}}\left[-\Lambda+\gamma\arctan\left(\frac{\Lambda}{\gamma}\right)\right],
\end{equation}
where $\gamma=\sqrt{-2\mu E}$ is the binding momentum. If we set the
$\Lambda\gg \gamma$,
\begin{equation}
    F_{1}\approx \frac{8\mu\pi}{(2\pi)^{3}}\left[-\Lambda+\gamma\frac{\pi}{2}\right].
\end{equation}
 We define $\omega=\sqrt{2\mu\delta}$. For $\Lambda\gg \omega$, $F_2$ reads
 \begin{equation}
    F_2 \approx \frac{8\mu\pi}{(2\pi)^{3}}\left[-\Lambda+\sqrt{\gamma^{2}+\omega^{2}}\frac{\pi}{2}\right].
 \end{equation}
Solving Eq.~\eqref{eq:cplse}, we obtain
\begin{eqnarray}
    t^{-1}&=& \left[\begin{array}{cc}
        \frac{v_{22}}{v_{11}v_{22}-v_{12}^{2}}-F_{1} & -\frac{v_{12}}{v_{11}v_{22}-v_{12}^{2}}\\
        -\frac{v_{12}}{v_{11}v_{22}-v_{12}^{2}} & \frac{v_{11}}{v_{11}v_{22}-v_{12}^{2}}-F_{2}
    \end{array}\right],
\end{eqnarray}
In order to discuss the cutoff-dependence, we introduce $a_{11}$,
$a_{12}$ and $a_{22}$ as the combination of $v_{ij}$ to make
\begin{eqnarray}
    t^{-1}&=& \left[\begin{array}{cc}
        a_{11}-F_{1} & a_{12}\\
        a_{12} & a_{22}-F_{2}
    \end{array}\right].~\label{eq:tinv}
\end{eqnarray}
The relations of $a_{ij}$ and $v_{ij}$ read
\begin{eqnarray}
    &&v_{11}=\frac{a_{22}}{a_{11}a_{22}-a_{12}^{2}},\quad v_{22}=\frac{a_{11}}{a_{11}a_{22}-a_{12}^{2}},\nonumber\\
    &&v_{12}=-\frac{a_{12}}{a_{11}a_{22}-a_{12}^{2}}.
\end{eqnarray}
In order to make the $t$-matrix cutoff-independent, the $a_{11}$ and
$a_{22}$ depend on $\Lambda$ as
\begin{eqnarray}
&&a_{11}(\Lambda)=a_{11}(\Lambda_{0})-\frac{\mu}{\pi^{2}}(\Lambda-\Lambda_{0}),\nonumber\\
&&a_{22}(\Lambda)=a_{22}(\Lambda_{0})-\frac{\mu}{\pi^{2}}(\Lambda-\Lambda_{0}).\label{eq:cutdepend}
\end{eqnarray}
The above relations make $(a_{11}-F_1)$ and $(a_{22}-F_2)$ cutoff
independent. $a_{12}$ itself is cutoff-independent.

\begin{figure*}
 \centering
\includegraphics[width=0.33\textwidth]{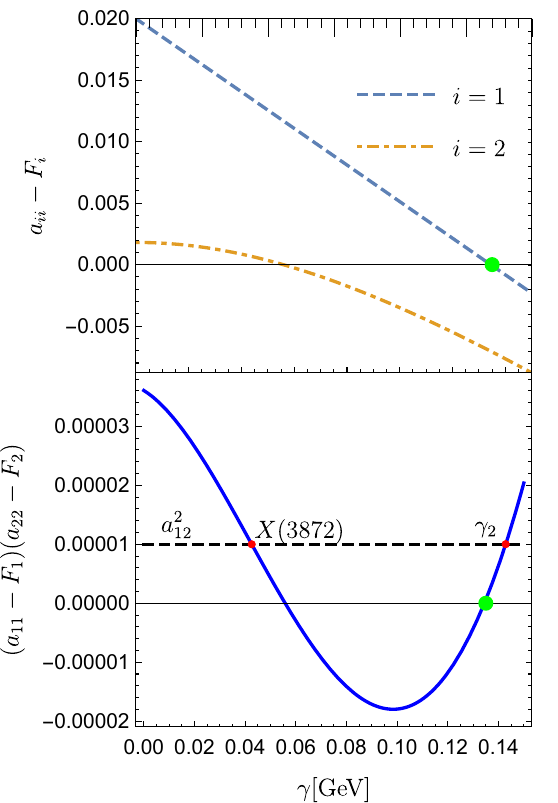}
\includegraphics[width=0.32\textwidth]{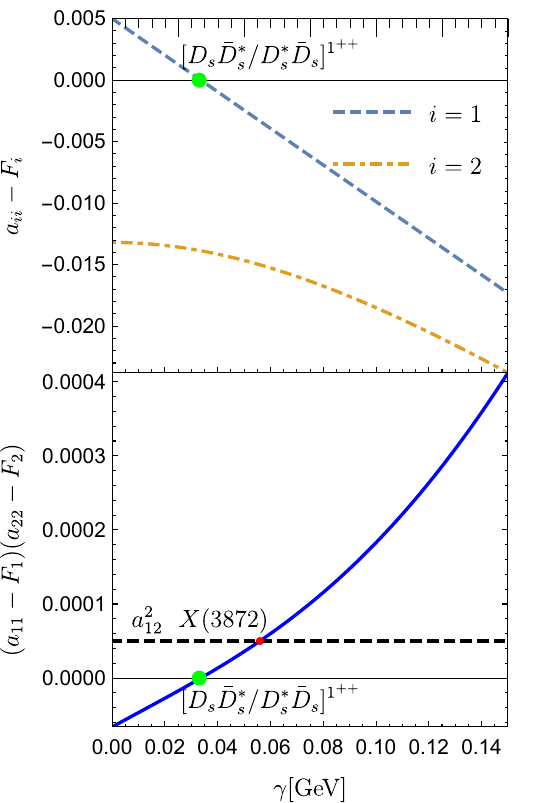}
\includegraphics[width=0.32\textwidth]{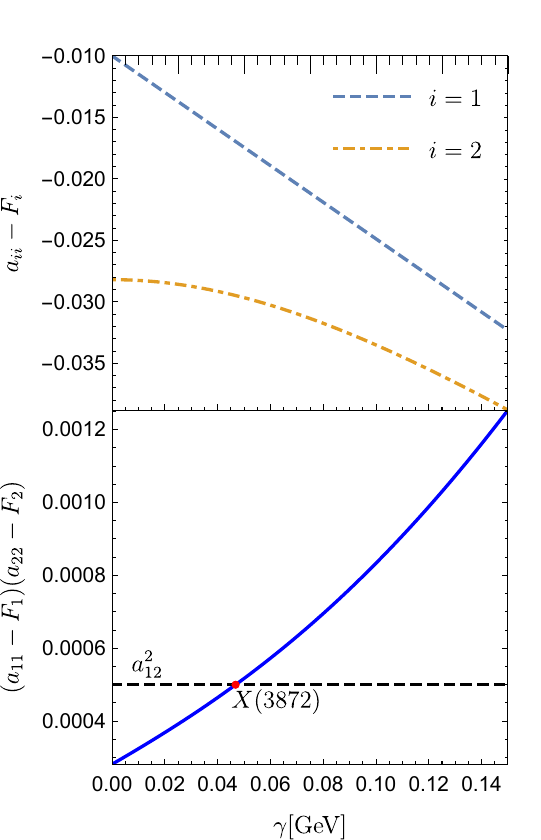}
\caption{The possible solutions of Eq.~\eqref{eq:det_a}. We use the
blue dashed line and orange dotdashed line to denote the
$a_{11}-F_1$ and $a_{22}-F_2$, respectively. Their product is
presented in blue solid line. The black dashed line represents the
$a_{12}^2$. The cross points of black dashed line and blue solid
line represent the binding solutions of Eq.~\eqref{eq:det_a}, which
are labeled by red points. We use the green points to denote the
solutions of $a-F(\gamma)=0$ (the same as $a_{11}-F_1(\gamma)=0$),
which represent the binding solutions of
$[D_s\bar{D}_s^*/D_s^*\bar{D}_s]^{1^{++}}$ system. The left
subfigure represents the case with $a_{11}-F_1(\gamma_1)>0$ and
$a_{22}-F_2(\gamma_1)>0$. The middle one and the right one represent
cases with $a_{11}-F_1(\gamma_1)<0$ and $a_{22}-F_2(\gamma_1)<0$. In
the middle subfigure, there is a binding solution for
$[D_s\bar{D}_s^*/D_s^*\bar{D}_s]^{1^{++}}$. In the right subfigure,
there does not exist $[D_s\bar{D}_s^*/D_s^*\bar{D}_s]^{1^{++}}$
bound state.}\label{fig:3case}
\end{figure*}

From Eq.~\eqref{eq:tinv}, the $t$-matrix reads
\begin{eqnarray}
t&=&\left[\begin{array}{cc}
    \frac{1}{(a_{11}-F_{1})-\frac{a_{12}^{2}}{a_{22}-F_{2}}} & \frac{-a_{12}}{(a_{11}-F_{1})(a_{22}-F_{2})-a_{12}^{2}}\\
    \frac{-a_{12}}{(a_{11}-F_{1})(a_{22}-F_{2})-a_{12}^{2}} & \frac{1}{(a_{22}-F_{2})-\frac{a_{12}^{2}}{a_{11}-F_{1}}}
\end{array}\right].
\end{eqnarray}
The pole of the $t$-matrix is obtained by
\begin{equation}
    [a_{11}-F_{1}(\gamma)][a_{22}-F_{2}(\gamma)]-a_{12}^{2}=0\label{eq:det_a},
\end{equation}
where
\begin{eqnarray}
    a_{11}-F_{1}(\gamma)&=&C-{\mu\over 2\pi}\gamma,\nonumber\\
    a_{22}-F_{2}(\gamma)&=&C-{\mu\over 2\pi}\sqrt{\gamma^2+\omega^2}.
\end{eqnarray}
$C$ is a constant. One of the solutions is
$\gamma_1=\gamma_{X(3872)}$. If $a_{11}-F_{1}(\gamma_1) > 0$ and
$a_{22}-F_{2}(\gamma_1) > 0$, there should exist the second binding
solution with $\gamma_2>\gamma_1$. As shown in the left plot in
Fig.~\ref{fig:3case}, $a_{11}-F_1(\gamma)$ and $a_{22}-F_2(\gamma)$
will decrease with $\gamma$ and the second binding solution
$\gamma_2$ appear with $a_{11}-F_1(\gamma_2)<0$ and
$a_{22}-F_2(\gamma_2)<0$. However, there is no such a state observed
in experiments. Thus, $a_{11}-F_{1}(\gamma_1) < 0$.

When we try to relate the $v_{11}$ in the coupled-channel formalism
(for $[\bar{D}^* D/\bar{D} D^*]^{1^{++}}$) to the $v$ in the
single-channel calculation (for the $[\bar{D}_s^* D_s/\bar{D}_s
D^*_s]^{1^{++}}$ system), we cannot expect
$v_{11}(\Lambda)=v(\Lambda)$ is always valid for all the cutoff
scale because the cutoff-dependence of $v_{11}$ is affected by the
second channel.  However, the relation of $a_{11}$ and $a$ reads
 \begin{equation}
    a_{11}(\Lambda)=a(\Lambda).\label{eq:acut}
\end{equation}
The validity of above relation is cutoff-independent. In the
single-channel calculation, $a$ depends on $\Lambda$ as
$a(\Lambda)=a(\Lambda_0)-{\mu \over \pi^2} (\Lambda-\Lambda_{0})$,
which is similar to Eq.~\eqref{eq:cutdepend}. Thus, the difference
of $a(\Lambda)$ and $a_{11}(\Lambda)$ is a cutoff-independent
constant. We will choose a $\Lambda_0$ to fix their difference. The
cutoff in Eqs.~\eqref{eq:cplse} and \eqref{eq:cpf} is a scale to
regulate the divergence in LSEs. The freedom and dynamics at the
larger scale than $\Lambda$ are absorbed into the cutoff-dependent
coupling constants $v_{ij}$. At a very large scale, the tiny
coupled-channel effect is absorbed into the coupling constants.
Thus, we can expect the $v= v_{11}$ at the large $\Lambda$ limit.
When we take $\Lambda_0 \gg  \sqrt{2\mu \delta}$ and $\Lambda_{0}\gg
{\pi^2\over \mu }a_{12}$, we have,
\begin{eqnarray}
   \frac{1}{v_{11}(\Lambda_0)}&=&a_{11}(\Lambda_0)-\frac{a_{12}^{2}}{a_{22}(\Lambda_0)} = a_{11}(\Lambda_0),\\
   \frac{1}{v_{11}(\Lambda_0)}&=& {1\over v (\Lambda_{0})}=a(\Lambda_{0}).
\end{eqnarray}
We can obtain Eq.~\eqref{eq:acut} from
$a_{11}(\Lambda_{0})=a(\Lambda_{0})$. This relation can also be
rewritten as
\begin{equation}
  \frac{1}{v_{11}(\Lambda)}+\frac{a_{12}^{2}}{a_{22}(\Lambda)}=  \frac{1}{v(\Lambda)}.
\end{equation}
 Thus, we relate the
interaction of $X(3872)$ in the coupled-channel formalism to the
$[\bar{D}_s^* D_s/\bar{D}_s D^*_s]^{1^{++}}$ interaction.

For the $[\bar{D}_s^* D_s/\bar{D}_s D^*_s]^{1^{++}}$ system,
$a-F(\gamma_1)=a_{11}-F_{11}(\gamma_1) < 0$ . Thus, there are two
possibilities for its binding solution as shown in the middle plot
and right plot in Fig.~\ref{fig:3case}, respectively. In the first
case, the possible binding solution of $[\bar{D}_s^* D_s/\bar{D}_s
D^*_s]^{1^{++}}$ appears with binding momentum $\gamma_{[\bar{D}_s^*
D_s]^{1^{++}}}$, where $\gamma_{[\bar{D}_s^*
D_s]^{1^{++}}}<\gamma_1$. In other words,  $[\bar{D}_s^*
D_s/\bar{D}_s D^*_s]^{1^{++}}$ is more shallowly bound state than
$X(3872)$, which indicates there is tiny parameter region for
$[\bar{D}_s^* D_s/\bar{D}_s D^*_s]^{1^{++}}$ to form bound state. It
is more likely that there is no binding solution for $[\bar{D}_s^*
D_s/\bar{D}_s D^*_s]^{1^{++}}$ system as shown in the right plot in
Fig.~\ref{fig:3case}. In both cases, the $[\bar{D}_s^* D_s/\bar{D}_s
D^*_s]^{1^{++}}$ interaction extracted from the coupled-channel
scheme is less attractive (more repulsive) than that from
single-channel calculation in Sec.~\ref{sec:dsds} (In
Appendix~\ref{app:pole_traj}, we will rule out the possibility
corresponding to middle subfigure in Fig.~\ref{fig:3case}.).
According to Fig.~\ref{fig:inter_order}, the less attractive (more
repulsive) $[\bar{D}_s^* D_s/\bar{D}_s D^*_s]^{1^{++}}$ interaction
is, the more attractive $[\bar{D}_s^* D_s/\bar{D}_s
D^*_s]^{1^{+-}}$, $[\bar{D}_s^{*}D_s^*]^{1^{+-}}$ and
$[\bar{D}_s^{*}D_s^*]^{0^{++}}$ channels are. Thus, the existences
of $[\bar{D}_s^* D_s/\bar{D}_s D^*_s]^{1^{+-}}$,
$[\bar{D}_s^{*}D_s^*]^{1^{+-}}$ and $[\bar{D}_s^{*}D_s^*]^{0^{++}}$
bound states will not change. One can notice that we do not specify
the mechanism of large isospin violation decays of $X(3872)$. Our
predictions are not affected by the underlying mechanism
qualitatively.

\section{Summary and Discussion}\label{sec:sum}

In this work, we prove the existence of the
$[\bar{D}_{s}^{*}D_{s}^{*}]^{0^{++}}$,
$[\bar{D}_{s}^{*}D_{s}/\bar{D}_{s}^{}D_{s}^*]^{1^{+-}}$, and
$[\bar{D}_{s}^{*}D_{s}^{*}]^{1^{+-}}$ bound states, which is the
consequence of two prerequisites in the $\SU3$ symmetry and HQSS.
The first prerequisite, the $X(3872)$ as a loosely molecular state
is supported by its mass and decay branching ratios. The observation
of $\chi_{c0}(3930)$~\cite{Aaij:2020hon,Aaij:2020ypa} and lattice
QCD calculation~\cite{Prelovsek:2020eiw} justify the second
prerequisite, the existence of $[\bar{D}_{s}D_{s}]^{0^{++}}$ bound
state. With these two prerequisites, there do exist the
$[\bar{D}_{s}^{*}D_{s}^{*}]^{0^{++}}$,
$[\bar{D}_{s}^{*}D_{s}/\bar{D}_{s}^{}D_{s}^*]^{1^{+-}}$, and
$[\bar{D}_{s}^{*}D_{s}^{*}]^{1^{+-}}$ bound states, which is the
natural consequence of $\SU3$ symmetry and HQSS. These three states
are all deeper bound states than the $[\bar{D}_{s}D_{s}]^{0^{++}}$
system, while the $[\bar{D}_{s}^{*}D_{s}^{*}]^{0^{++}}$ is the
deepest one.

As shown in Sec.~\ref{sec:symmtry}, the strange quark destroys the
$\SU3$ symmetry for the di-meson systems, which suppresses the
mixture of the $\bar{D}_s^{(*)}D_s^{(*)}$ and $\bar{D}^{(*)}D^{(*)}$
systems. However, the isospin symmetry is still a very good
approximation for the di-meson systems, which is supported by the
observation of the almost degenerate $Z_c(3900)/Z_c(4020)$ isospin
triplet~\cite{Ablikim:2015gda,Ablikim:2015tbp,Ablikim:2015vvn,Zyla:2020zbs}.
The neutral states are the half-and-half mixture of the
$\bar{D}^{(*)0}D^{(*)0}$ and ${D}^{(*)-}D^{(*)+}$. Compared with
$Z_c(3900)/Z_c(4020)$ states, the $X(3872)$ with a large isospin
breaking effect is very unusual even in the exotic hadron family.
The peculiarities of the $X(3872)$ might stem from either the
accidental fine-tuning of the $\bar{D}^{*}D/\bar{D} D^{*}$
interaction or the accidental fine-tuning of $P$-wave charmonium
state to the $\bar{D}^{*0}D^0/\bar{D}^0 D^{*0}$
threshold~\cite{Braaten:2003he}. Our proof is established upon the
first fine-tuning. If the $X(3872)$ is a consequence of the second
fine-tuning, the $P$-wave charmonium state coinciding with the
$\bar{D}^{*0}D^0/\bar{D}^0 D^{*0}$ threshold, the absence of the
$[D^{*-}D^+/D^- D^{*+}]^{1^{++}}$ di-meson state would imply its
weakly attractive or repulsive hadronic interaction. The unbound
$[D^{*-}D^+/D^- D^{*+}]^{1^{++}}$ system together with the existence
of the $[\bar{D}_{s}D_{s}]^{0^{++}}$ bound state would induce to the
same consequence in the $\SU3$ symmetry and HQSS limit.

Among the six $S$-wave $\bar{D}_s^{(*)}D_s^{(*)}$ systems, we have
discussed four of them. The remaining $[\bar{D}_s^*D_s/\bar{D}_s
D^*]^{1^{++}}$ and $[\bar{D}^*_s D^*_s]^{2^{++}}$ have the same
interaction as the $X(3872)$ in the single-channel analysis and have
the less attractive (more repulsive) interaction in the
coupled-channel interaction. However, we do not expect these two
systems have the similar fine-tuning mechanism as the $X(3872)$.
Therefore, there may not exist the di-meson states in these two
channels. In Ref.~\cite{Molina:2009ct}, the $X(4140)$ was
dynamically generated as $D_s^*\bar{D}_s^*$ molecule with
$J^{PC}=2^{++}$.

In Ref.~\cite{HidalgoDuque:2012pq}, the authors treated $X(3872)$,
$X(3915)$, and $Y(4140)$ as $[D^*\bar{D}/\bar{D}^*D]^{1^{++}}$,
$[D^*\bar{D}^*]^{0^{++}}$, $[D_s^*\bar{D}_s^*]^{0^{++}}$ molecules,
respectively and predicted the spectrum of heavy meson molecules. We
compare their $D_s^{(*)}\bar{D}_s^{(*)}$ results with our
predictions in Table~\ref{tab:be}. Their results agree with our
predictions, though we choose the different states as the inputs.
Meanwhile, they neither obtained the $[\bar{D}_s^*D_s/\bar{D}_s
D^*]^{1^{++}}$ and $[\bar{D}^*_s D^*_s]^{2^{++}}$ molecular states.
In our analysis, we do not choose $X(3915)$ as the input due to its
controversial quantum number and nature. The experimental analysis
prefer its $J^{PC}=0^{++}$~\cite{Lees:2012xs}. However, a reanalysis
presented in Ref.~\cite{Zhou:2015uva} showed that a $J^{PC}=2^{++}$
is also possible. Apart from the molecular candidate of
$[D^*\bar{D}^*]^{0^{++}}$~\cite{HidalgoDuque:2012pq}, the $X(3915)$
was also interpreted as the $[D_s\bar{D}_s]^{0^{++}}$ bound
states~\cite{Li:2015iga}, $c\bar{c}s\bar{s}$ tetraquark
state~\cite{Lebed:2016yvr} and $P$-wave
charmonium~\cite{Liu:2009fe}.

If $X(3872)$ is not a bound state but a threshold effect with a
virtual pole near the threshold, we can infer that the
$\bar{D}^{*0}D^0/\bar{D}^0 D^{*0}$ interaction is either repulsive
or not as attractive as we expected in Sections~\ref{sec:dsds} and
\ref{sec:cc}. In this case, our predictions will not change
qualitatively. According to Eq.~\eqref{eq:ratio}, the less
attractive (more repulsive) the $[\bar{D}^{*0}D^0/\bar{D}^0
D^{*0}]^{1^{++}}$ interaction is, the more attractive the
$[\bar{D}_s^* D_s/\bar{D}_s D^*_s]^{1^{+-}}$,
$[\bar{D}_s^{*}D_s^*]^{1^{+-}}$ and $[\bar{D}_s^{*}D_s^*]^{0^{++}}$
channels are.

We could estimate the isospin breaking effect stemming from the
meson-exchange interaction. The mass difference between the pion and
other pseudoscalar mesons will give rise to the main part of the
isospin breaking effect (For the $\mathtt{PP}$ systems, the
pseudoscalar-meson-exchange interaction is forbidden, for which the
meson-exchange interactions arise from either the
vector-meson-exchange effect or coupled-channel effect.  ). For the
$\bar{D}_0^{(*)}D_0^{(*)}$ systems, the $\pi$ and $\eta$ are
exchanged. For the $\bar{D}_s^{(*)}D_s^{(*)}$ system, only the
$\eta$-exchange interaction is allowed. The interactions read
    \begin{eqnarray}
        {\cal V}_{\bar{u}u}&\sim&\frac{1}{6}\frac{1}{q^{2}-m_{\pi}^{2}}+\frac{1}{2}\frac{1}{q^{2}-m_{\eta}^{2}},\nonumber \\{\cal V}_{\bar{s}s}&\sim&\frac{2}{3}\frac{1}{q^{2}-m_{\eta}^{2}}.
    \end{eqnarray}
    If we integrate out the $\pi$ and $\eta$ degrees of freedom and match with the pion-less EFT, we roughly have ${\cal V}_{\bar{u}u}/{\cal V}_{\bar{s}s}\approx 4.6$. In other words, including the isospin breaking effect will reduce the $[\bar{D}_{s}^{*}D_{s}/\bar{D}_{s}^{}D_{s}^*]^{1^{++}}$ interaction by one order and make it less attractive. According to Eq.~\eqref{eq:ratio}, the less attractive $[\bar{D}_{s}^{*}D_{s}/\bar{D}_{s}^{}D_{s}^*]^{1^{++}}$ will lead to more attractive $[\bar{D}_s^* D_s/\bar{D}_s
    D^*_s]^{1^{+-}}$, $[\bar{D}_s^{*}D_s^*]^{1^{+-}}$ and
    $[\bar{D}_s^{*}D_s^*]^{0^{++}}$ interactions. Thus, our prediction will not change qualitatively by the isospin breaking effect.

In the lattice QCD simulation~\cite{Prelovsek:2020eiw}, there exists
the $0^{++}$ $\bar{D}_sD_s$ bound state in the one-channel
approximation. After considering the coupled-channel effect of
$\bar{D}{D}$, there is an indication for a narrow $0^{++}$ resonance
just below the $\bar{D}_s{D}_s$ threshold with a large coupling to
$\bar{D}_s{D}_s$ and a very small coupling to $\bar{D}D$. The
suppression of coupled-channel effect between $\bar{D}_s D_s$ and
$\bar{D}D$ is in agreement with our analysis in
Sec.~\ref{sec:symmtry}. Apart from the $\bar{D}D$ channel, the
$\bar{D}_s D_s$ states might be affected by the $\bar{D}^*D^*$
channel~\cite{Molina:2009ct,Gamermann:2006nm}.

We hope the future analyses in experiments can search for the
$[\bar{D}_{s}^{*}D_{s}^{*}]^{0^{++}}$,
$[\bar{D}_{s}^{*}D_{s}/\bar{D}_{s}^{}D_{s}^*]^{1^{+-}}$, and
$[\bar{D}_{s}^{*}D_{s}^{*}]^{1^{+-}}$ bound states in the $B\to
D_{(s)}^{(*)}\bar{D}^{(*)}_{(s)}h$ processes ($h$ denotes the light
hadrons). The $[\bar{D}_{s}^{*}D_{s}^{*}]^{0^{++}}$ bound state is
also expected to be observed in the $J/\psi \phi$ final state in the
$B\to J/\psi \phi K$ decay~\cite{Dong:2020hxe}. In fact, we notice
there seems to exist some excess near $4200$ MeV in the $J/\psi
\phi$ invariant mass spectrum in the previous analysis of LHCb
Collaboration~\cite{Aaij:2016iza,Aaij:2016nsc}.

\vspace{0.5cm}

\begin{acknowledgements}
We are grateful to Evgeny Epelbaum for a careful reading of the
manuscript and helpful comments. We thank Eulogio Oset for helpful
comments. This project is supported by the National Natural Science
Foundation of China under Grants No. 11975033 and 12070131001. This
project is also funded by the Deutsche Forschungsgemeinschaft (DFG,
German Research Foundation) -Project-ID 196253076 -TRR 110.
\end{acknowledgements}

\begin{appendix}
    \section{Pole trajectories in the coupled-channel formalism}~\label{app:pole_traj}
In the coupled-channel formalism, the existence of the pole
corresponding to $X(3872)$ will reduce the three parameters,
$a_{11}$, $a_{22}$ and $a_{12}$, to two independent ones in
Eq.~\eqref{eq:det_a}. If we presume the coupling constants satisfy
the $\SU3$ symmetry ($a_{11}=a_{22}\equiv a_{ii}$), only one free
parameter $a_{ii}$ is left. The $a_{12}$ reads,
\begin{equation}
    a_{12}^2=[a_{ii}-F_1(\gamma_{X(3872)})][a_{ii}-F_2(\gamma_{X(3872)})].\label{eq:constrain1}
\end{equation}
Meanwhile, the non-existence of the other $1^{++}$ states around the
$\bar{D}D^*$ threshold except $X(3872)$, will also constrain the
range of the only parameter $a_{ii}$. By ruling out the existence of
the second bound state (see the left subfigure in
Fig.~\ref{fig:3case}),  we obtain
 \begin{equation}
    [a_{ii}-F_1(\gamma_{X(3872)})]<[a_{ii}-F_2(\gamma_{X(3872)})]<0.\label{eq:constrain2}
 \end{equation}
In this section, we will explore the constraint from the
non-existence of the near-threshold poles in the complex energy
plane.

Under the constraints of Eqs.~\eqref{eq:constrain1} and
\eqref{eq:constrain2}, we decrease the parameter $a_{ii}$
(equivalently increase $a_{12}^2$) and obtain the trajectories of
the poles in the four Riemann sheets. We use the
$\text{RS}_{\pm\pm}$ to denote the Riemann sheets, where two
subscripts stand for the sign of $\text{Im}k_1$ and $\text{Im}k_2$,
respectively. $k_i$ is the momentum in the center of mass frame for
the $i$th channel. In order to demonstrate the trajectories of the
poles conveniently, we fix the binding energy of the $X(3872)$ as
$1$ MeV. The slight variation of the bind energy will not change the
results qualitatively. We start from $a_{12}^2=0$, which corresponds
to the vanishing coupled-channel effect. There exist poles in the
real axises of $\text{RS}_{+-}$ and $\text{RS}_{-+}$. The pole in
the $\text{RS}_{-+}$ represents the bound solution corresponding to
the charged $D^*\bar{D}/\bar{D}D^*$ systems. In the range (a), the
pole in the lower plane of $\text{RS}_{-+}$ corresponds to a
resonance state. Its width increases with the $a_{12}^2$ and finally
achieves its maximum at the threshold of $D^+D^{*-}$. At the end of
the range (a), the pole in $\text{RS}_{+-}$ disappears at the
$D^0\bar{D}^{*0}$ threshold. The pole in the $\text{RS}_{--}$
appears in the range (b) and (c) and moves along the real axis to be
away from the $D\bar{D}^*$ threshold. The poles in the
$\text{RS}_{-+}$ will finally disappear in the real axis at the end
of range (b), where the poles in $\text{RS}_{+-}$ will appear at the
beginning of range (c). Finally, with the increasing of $a_{12}^2$,
the poles except the one corresponding to $X(3872)$ are away from
the $D\bar{D}^*$ threshold. In the other words, decreasing $a_{ii}$
(equivalently increasing $a_{12}^2$) would achieve the range of
parameters corresponding to the experimental observations around the
$D\bar{D}^*$ threshold.

Since the non-existence of the other resonance states except
$X(3872)$ constrains a relative large $a_{12}^2$ (the large
coupled-channel effect), the case in the middle subfigure of
Fig.~\ref{fig:3case} could be ruled out. Thus, there does not exist
the $[D_s\bar{D}_s^*/D_s^*\bar{D}_s]^{1^{++}}$ bound state.

    \begin{figure}[!htp]
        \centering
        \includegraphics[width=0.45\textwidth]{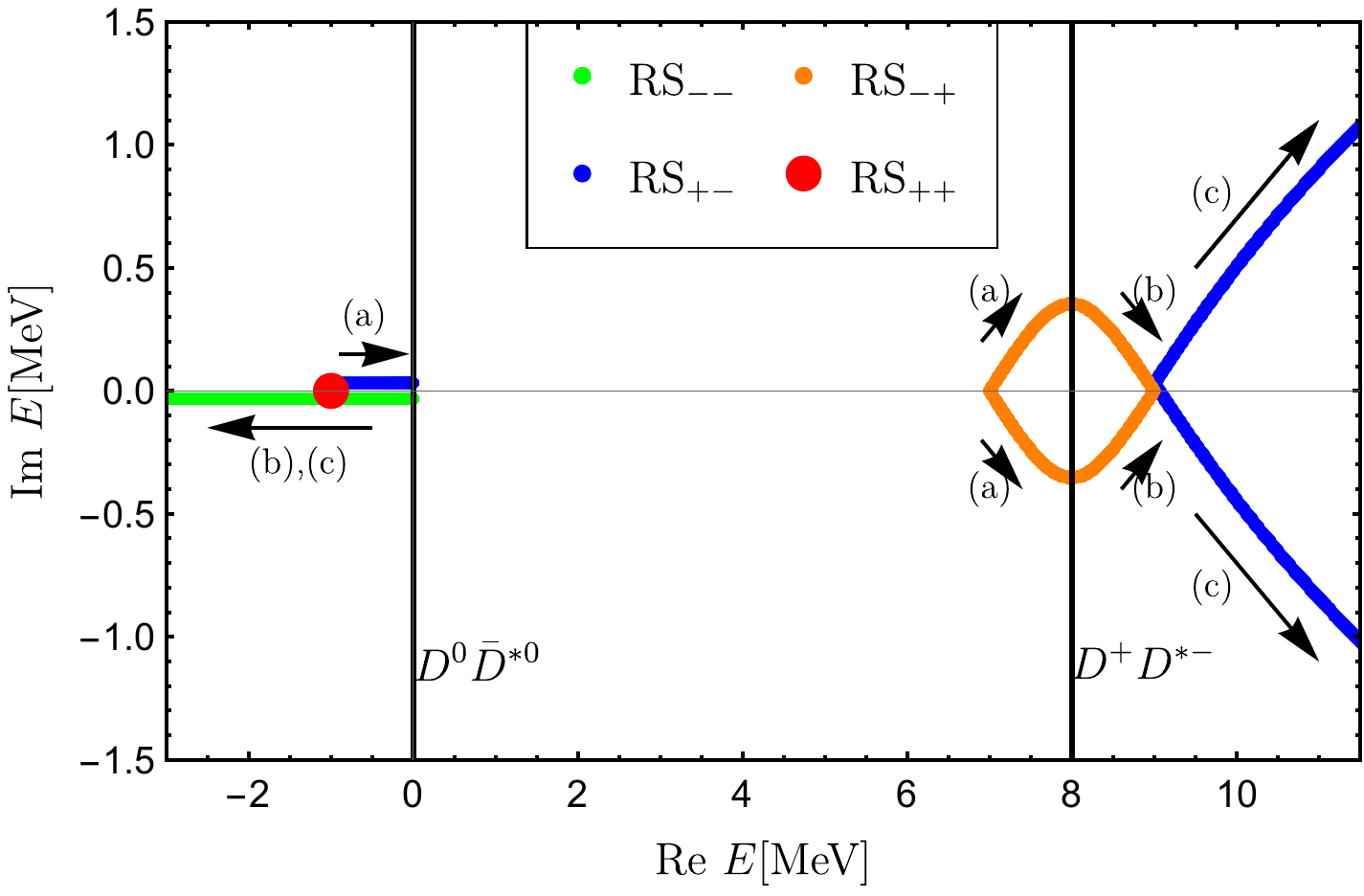}
        \caption{The trajectories of poles in the four Riemann sheets with the decreasing of the parameter $a_{ii}$ (equivalently increasing $a_{12}^2$). We use arrow to denote the direction of trajectories of poles. The (a), (b), (c) are three ranges of $a_{12}^2$ in ascending order. We fix the binding energy of $X(3872)$ as $1$ MeV. }\label{fig:pole}
    \end{figure}
\end{appendix}

%\bibliographystyle{apsrmp4-1.bst}
%\bibliography{C:/Users/dreamway/OneDrive/Academic/Documents/paper}
\bibliography{ref}

\begin{thebibliography}{67}%
\makeatletter
\providecommand \@ifxundefined [1]{%
 \@ifx{#1\undefined}
}%
\providecommand \@ifnum [1]{%
 \ifnum #1\expandafter \@firstoftwo
 \else \expandafter \@secondoftwo
 \fi
}%
\providecommand \@ifx [1]{%
 \ifx #1\expandafter \@firstoftwo
 \else \expandafter \@secondoftwo
 \fi
}%
\providecommand \natexlab [1]{#1}%
\providecommand \enquote  [1]{``#1''}%
\providecommand \bibnamefont  [1]{#1}%
\providecommand \bibfnamefont [1]{#1}%
\providecommand \citenamefont [1]{#1}%
\providecommand \href@noop [0]{\@secondoftwo}%
\providecommand \href [0]{\begingroup \@sanitize@url \@href}%
\providecommand \@href[1]{\@@startlink{#1}\@@href}%
\providecommand \@@href[1]{\endgroup#1\@@endlink}%
\providecommand \@sanitize@url [0]{\catcode `\\12\catcode `\$12\catcode
  `\&12\catcode `\#12\catcode `\^12\catcode `\_12\catcode `\%12\relax}%
\providecommand \@@startlink[1]{}%
\providecommand \@@endlink[0]{}%
\providecommand \url  [0]{\begingroup\@sanitize@url \@url }%
\providecommand \@url [1]{\endgroup\@href {#1}{\urlprefix }}%
\providecommand \urlprefix  [0]{URL }%
\providecommand \Eprint [0]{\href }%
\providecommand \doibase [0]{http://dx.doi.org/}%
\providecommand \selectlanguage [0]{\@gobble}%
\providecommand \bibinfo  [0]{\@secondoftwo}%
\providecommand \bibfield  [0]{\@secondoftwo}%
\providecommand \translation [1]{[#1]}%
\providecommand \BibitemOpen [0]{}%
\providecommand \bibitemStop [0]{}%
\providecommand \bibitemNoStop [0]{.\EOS\space}%
\providecommand \EOS [0]{\spacefactor3000\relax}%
\providecommand \BibitemShut  [1]{\csname bibitem#1\endcsname}%
\let\auto@bib@innerbib\@empty
%</preamble>
\bibitem [{\citenamefont {Prelovsek}\ \emph {et~al.}(2020)\citenamefont
  {Prelovsek}, \citenamefont {Collins}, \citenamefont {Mohler}, \citenamefont
  {Padmanath},\ and\ \citenamefont {Piemonte}}]{Prelovsek:2020eiw}%
  \BibitemOpen
  \bibfield  {author} {\bibinfo {author} {\bibfnamefont {S.}~\bibnamefont
  {Prelovsek}}, \bibinfo {author} {\bibfnamefont {S.}~\bibnamefont {Collins}},
  \bibinfo {author} {\bibfnamefont {D.}~\bibnamefont {Mohler}}, \bibinfo
  {author} {\bibfnamefont {M.}~\bibnamefont {Padmanath}}, \ and\ \bibinfo
  {author} {\bibfnamefont {S.}~\bibnamefont {Piemonte}},\ }\href@noop {} {\
  (\bibinfo {year} {2020})},\ \Eprint {http://arxiv.org/abs/2011.02542}
  {arXiv:2011.02542 [hep-lat]} \BibitemShut {NoStop}%
\bibitem [{\citenamefont {Aaij}\ \emph
  {et~al.}(2020{\natexlab{a}})\citenamefont {Aaij} \emph
  {et~al.}}]{Aaij:2020hon}%
  \BibitemOpen
  \bibfield  {author} {\bibinfo {author} {\bibfnamefont {R.}~\bibnamefont
  {Aaij}} \emph {et~al.} (\bibinfo {collaboration} {LHCb}),\ }\href {\doibase
  10.1103/PhysRevLett.125.242001} {\bibfield  {journal} {\bibinfo  {journal}
  {Phys. Rev. Lett.}\ }\textbf {\bibinfo {volume} {125}},\ \bibinfo {pages}
  {242001} (\bibinfo {year} {2020}{\natexlab{a}})},\ \Eprint
  {http://arxiv.org/abs/2009.00025} {arXiv:2009.00025 [hep-ex]} \BibitemShut
  {NoStop}%
\bibitem [{\citenamefont {Aaij}\ \emph
  {et~al.}(2020{\natexlab{b}})\citenamefont {Aaij} \emph
  {et~al.}}]{Aaij:2020ypa}%
  \BibitemOpen
  \bibfield  {author} {\bibinfo {author} {\bibfnamefont {R.}~\bibnamefont
  {Aaij}} \emph {et~al.} (\bibinfo {collaboration} {LHCb}),\ }\href {\doibase
  10.1103/PhysRevD.102.112003} {\bibfield  {journal} {\bibinfo  {journal}
  {Phys. Rev. D}\ }\textbf {\bibinfo {volume} {102}},\ \bibinfo {pages}
  {112003} (\bibinfo {year} {2020}{\natexlab{b}})},\ \Eprint
  {http://arxiv.org/abs/2009.00026} {arXiv:2009.00026 [hep-ex]} \BibitemShut
  {NoStop}%
\bibitem [{\citenamefont {Barnes}\ \emph {et~al.}(1964)\citenamefont {Barnes}
  \emph {et~al.}}]{Barnes:1964pd}%
  \BibitemOpen
  \bibfield  {author} {\bibinfo {author} {\bibfnamefont {V.~E.}\ \bibnamefont
  {Barnes}} \emph {et~al.},\ }\href {\doibase 10.1103/PhysRevLett.12.204}
  {\bibfield  {journal} {\bibinfo  {journal} {Phys. Rev. Lett.}\ }\textbf
  {\bibinfo {volume} {12}},\ \bibinfo {pages} {204} (\bibinfo {year}
  {1964})}\BibitemShut {NoStop}%
\bibitem [{\citenamefont {Choi}\ \emph {et~al.}(2003)\citenamefont {Choi} \emph
  {et~al.}}]{Choi:2003ue}%
  \BibitemOpen
  \bibfield  {author} {\bibinfo {author} {\bibfnamefont {S.~K.}\ \bibnamefont
  {Choi}} \emph {et~al.} (\bibinfo {collaboration} {Belle}),\ }\href {\doibase
  10.1103/PhysRevLett.91.262001} {\bibfield  {journal} {\bibinfo  {journal}
  {Phys. Rev. Lett.}\ }\textbf {\bibinfo {volume} {91}},\ \bibinfo {pages}
  {262001} (\bibinfo {year} {2003})},\ \Eprint
  {http://arxiv.org/abs/hep-ex/0309032} {arXiv:hep-ex/0309032} \BibitemShut
  {NoStop}%
\bibitem [{\citenamefont {Brambilla}\ \emph {et~al.}(2020)\citenamefont
  {Brambilla}, \citenamefont {Eidelman}, \citenamefont {Hanhart}, \citenamefont
  {Nefediev}, \citenamefont {Shen}, \citenamefont {Thomas}, \citenamefont
  {Vairo},\ and\ \citenamefont {Yuan}}]{Brambilla:2019esw}%
  \BibitemOpen
  \bibfield  {author} {\bibinfo {author} {\bibfnamefont {N.}~\bibnamefont
  {Brambilla}}, \bibinfo {author} {\bibfnamefont {S.}~\bibnamefont {Eidelman}},
  \bibinfo {author} {\bibfnamefont {C.}~\bibnamefont {Hanhart}}, \bibinfo
  {author} {\bibfnamefont {A.}~\bibnamefont {Nefediev}}, \bibinfo {author}
  {\bibfnamefont {C.-P.}\ \bibnamefont {Shen}}, \bibinfo {author}
  {\bibfnamefont {C.~E.}\ \bibnamefont {Thomas}}, \bibinfo {author}
  {\bibfnamefont {A.}~\bibnamefont {Vairo}}, \ and\ \bibinfo {author}
  {\bibfnamefont {C.-Z.}\ \bibnamefont {Yuan}},\ }\href {\doibase
  10.1016/j.physrep.2020.05.001} {\bibfield  {journal} {\bibinfo  {journal}
  {Phys. Rept.}\ }\textbf {\bibinfo {volume} {873}},\ \bibinfo {pages} {1}
  (\bibinfo {year} {2020})},\ \Eprint {http://arxiv.org/abs/1907.07583}
  {arXiv:1907.07583 [hep-ex]} \BibitemShut {NoStop}%
\bibitem [{\citenamefont {Liu}\ \emph {et~al.}(2019)\citenamefont {Liu},
  \citenamefont {Chen}, \citenamefont {Chen}, \citenamefont {Liu},\ and\
  \citenamefont {Zhu}}]{Liu:2019zoy}%
  \BibitemOpen
  \bibfield  {author} {\bibinfo {author} {\bibfnamefont {Y.-R.}\ \bibnamefont
  {Liu}}, \bibinfo {author} {\bibfnamefont {H.-X.}\ \bibnamefont {Chen}},
  \bibinfo {author} {\bibfnamefont {W.}~\bibnamefont {Chen}}, \bibinfo {author}
  {\bibfnamefont {X.}~\bibnamefont {Liu}}, \ and\ \bibinfo {author}
  {\bibfnamefont {S.-L.}\ \bibnamefont {Zhu}},\ }\href {\doibase
  10.1016/j.ppnp.2019.04.003} {\bibfield  {journal} {\bibinfo  {journal} {Prog.
  Part. Nucl. Phys.}\ }\textbf {\bibinfo {volume} {107}},\ \bibinfo {pages}
  {237} (\bibinfo {year} {2019})},\ \Eprint {http://arxiv.org/abs/1903.11976}
  {arXiv:1903.11976 [hep-ph]} \BibitemShut {NoStop}%
\bibitem [{\citenamefont {Guo}\ \emph {et~al.}(2018)\citenamefont {Guo},
  \citenamefont {Hanhart}, \citenamefont {Mei\ss{}ner}, \citenamefont {Wang},
  \citenamefont {Zhao},\ and\ \citenamefont {Zou}}]{Guo:2017jvc}%
  \BibitemOpen
  \bibfield  {author} {\bibinfo {author} {\bibfnamefont {F.-K.}\ \bibnamefont
  {Guo}}, \bibinfo {author} {\bibfnamefont {C.}~\bibnamefont {Hanhart}},
  \bibinfo {author} {\bibfnamefont {U.-G.}\ \bibnamefont {Mei\ss{}ner}},
  \bibinfo {author} {\bibfnamefont {Q.}~\bibnamefont {Wang}}, \bibinfo {author}
  {\bibfnamefont {Q.}~\bibnamefont {Zhao}}, \ and\ \bibinfo {author}
  {\bibfnamefont {B.-S.}\ \bibnamefont {Zou}},\ }\href {\doibase
  10.1103/RevModPhys.90.015004} {\bibfield  {journal} {\bibinfo  {journal}
  {Rev. Mod. Phys.}\ }\textbf {\bibinfo {volume} {90}},\ \bibinfo {pages}
  {015004} (\bibinfo {year} {2018})},\ \Eprint
  {http://arxiv.org/abs/1705.00141} {arXiv:1705.00141 [hep-ph]} \BibitemShut
  {NoStop}%
\bibitem [{\citenamefont {Olsen}\ \emph {et~al.}(2018)\citenamefont {Olsen},
  \citenamefont {Skwarnicki},\ and\ \citenamefont {Zieminska}}]{Olsen:2017bmm}%
  \BibitemOpen
  \bibfield  {author} {\bibinfo {author} {\bibfnamefont {S.~L.}\ \bibnamefont
  {Olsen}}, \bibinfo {author} {\bibfnamefont {T.}~\bibnamefont {Skwarnicki}}, \
  and\ \bibinfo {author} {\bibfnamefont {D.}~\bibnamefont {Zieminska}},\ }\href
  {\doibase 10.1103/RevModPhys.90.015003} {\bibfield  {journal} {\bibinfo
  {journal} {Rev. Mod. Phys.}\ }\textbf {\bibinfo {volume} {90}},\ \bibinfo
  {pages} {015003} (\bibinfo {year} {2018})},\ \Eprint
  {http://arxiv.org/abs/1708.04012} {arXiv:1708.04012 [hep-ph]} \BibitemShut
  {NoStop}%
\bibitem [{\citenamefont {Chen}\ \emph {et~al.}(2016)\citenamefont {Chen},
  \citenamefont {Chen}, \citenamefont {Liu},\ and\ \citenamefont
  {Zhu}}]{Chen:2016qju}%
  \BibitemOpen
  \bibfield  {author} {\bibinfo {author} {\bibfnamefont {H.-X.}\ \bibnamefont
  {Chen}}, \bibinfo {author} {\bibfnamefont {W.}~\bibnamefont {Chen}}, \bibinfo
  {author} {\bibfnamefont {X.}~\bibnamefont {Liu}}, \ and\ \bibinfo {author}
  {\bibfnamefont {S.-L.}\ \bibnamefont {Zhu}},\ }\href {\doibase
  10.1016/j.physrep.2016.05.004} {\bibfield  {journal} {\bibinfo  {journal}
  {Phys. Rept.}\ }\textbf {\bibinfo {volume} {639}},\ \bibinfo {pages} {1}
  (\bibinfo {year} {2016})},\ \Eprint {http://arxiv.org/abs/1601.02092}
  {arXiv:1601.02092 [hep-ph]} \BibitemShut {NoStop}%
\bibitem [{\citenamefont {Esposito}\ \emph {et~al.}(2017)\citenamefont
  {Esposito}, \citenamefont {Pilloni},\ and\ \citenamefont
  {Polosa}}]{Esposito:2016noz}%
  \BibitemOpen
  \bibfield  {author} {\bibinfo {author} {\bibfnamefont {A.}~\bibnamefont
  {Esposito}}, \bibinfo {author} {\bibfnamefont {A.}~\bibnamefont {Pilloni}}, \
  and\ \bibinfo {author} {\bibfnamefont {A.~D.}\ \bibnamefont {Polosa}},\
  }\href {\doibase 10.1016/j.physrep.2016.11.002} {\bibfield  {journal}
  {\bibinfo  {journal} {Phys. Rept.}\ }\textbf {\bibinfo {volume} {668}},\
  \bibinfo {pages} {1} (\bibinfo {year} {2017})},\ \Eprint
  {http://arxiv.org/abs/1611.07920} {arXiv:1611.07920 [hep-ph]} \BibitemShut
  {NoStop}%
\bibitem [{\citenamefont {Bondar}\ \emph {et~al.}(2012)\citenamefont {Bondar}
  \emph {et~al.}}]{Belle:2011aa}%
  \BibitemOpen
  \bibfield  {author} {\bibinfo {author} {\bibfnamefont {A.}~\bibnamefont
  {Bondar}} \emph {et~al.} (\bibinfo {collaboration} {Belle}),\ }\href
  {\doibase 10.1103/PhysRevLett.108.122001} {\bibfield  {journal} {\bibinfo
  {journal} {Phys. Rev. Lett.}\ }\textbf {\bibinfo {volume} {108}},\ \bibinfo
  {pages} {122001} (\bibinfo {year} {2012})},\ \Eprint
  {http://arxiv.org/abs/1110.2251} {arXiv:1110.2251 [hep-ex]} \BibitemShut
  {NoStop}%
\bibitem [{\citenamefont {Ablikim}\ \emph {et~al.}(2013)\citenamefont {Ablikim}
  \emph {et~al.}}]{Ablikim:2013wzq}%
  \BibitemOpen
  \bibfield  {author} {\bibinfo {author} {\bibfnamefont {M.}~\bibnamefont
  {Ablikim}} \emph {et~al.} (\bibinfo {collaboration} {BESIII}),\ }\href
  {\doibase 10.1103/PhysRevLett.111.242001} {\bibfield  {journal} {\bibinfo
  {journal} {Phys. Rev. Lett.}\ }\textbf {\bibinfo {volume} {111}},\ \bibinfo
  {pages} {242001} (\bibinfo {year} {2013})},\ \Eprint
  {http://arxiv.org/abs/1309.1896} {arXiv:1309.1896 [hep-ex]} \BibitemShut
  {NoStop}%
\bibitem [{\citenamefont {Ablikim}\ \emph {et~al.}(2014)\citenamefont {Ablikim}
  \emph {et~al.}}]{Ablikim:2013emm}%
  \BibitemOpen
  \bibfield  {author} {\bibinfo {author} {\bibfnamefont {M.}~\bibnamefont
  {Ablikim}} \emph {et~al.} (\bibinfo {collaboration} {BESIII}),\ }\href
  {\doibase 10.1103/PhysRevLett.112.132001} {\bibfield  {journal} {\bibinfo
  {journal} {Phys. Rev. Lett.}\ }\textbf {\bibinfo {volume} {112}},\ \bibinfo
  {pages} {132001} (\bibinfo {year} {2014})},\ \Eprint
  {http://arxiv.org/abs/1308.2760} {arXiv:1308.2760 [hep-ex]} \BibitemShut
  {NoStop}%
\bibitem [{\citenamefont {Aaij}\ \emph {et~al.}(2015)\citenamefont {Aaij} \emph
  {et~al.}}]{Aaij:2015tga}%
  \BibitemOpen
  \bibfield  {author} {\bibinfo {author} {\bibfnamefont {R.}~\bibnamefont
  {Aaij}} \emph {et~al.} (\bibinfo {collaboration} {LHCb}),\ }\href {\doibase
  10.1103/PhysRevLett.115.072001} {\bibfield  {journal} {\bibinfo  {journal}
  {Phys. Rev. Lett.}\ }\textbf {\bibinfo {volume} {115}},\ \bibinfo {pages}
  {072001} (\bibinfo {year} {2015})},\ \Eprint
  {http://arxiv.org/abs/1507.03414} {arXiv:1507.03414 [hep-ex]} \BibitemShut
  {NoStop}%
\bibitem [{\citenamefont {Aaij}\ \emph {et~al.}(2019)\citenamefont {Aaij} \emph
  {et~al.}}]{Aaij:2019vzc}%
  \BibitemOpen
  \bibfield  {author} {\bibinfo {author} {\bibfnamefont {R.}~\bibnamefont
  {Aaij}} \emph {et~al.} (\bibinfo {collaboration} {LHCb}),\ }\href {\doibase
  10.1103/PhysRevLett.122.222001} {\bibfield  {journal} {\bibinfo  {journal}
  {Phys. Rev. Lett.}\ }\textbf {\bibinfo {volume} {122}},\ \bibinfo {pages}
  {222001} (\bibinfo {year} {2019})},\ \Eprint
  {http://arxiv.org/abs/1904.03947} {arXiv:1904.03947 [hep-ex]} \BibitemShut
  {NoStop}%
\bibitem [{\citenamefont {Aaij}\ \emph
  {et~al.}(2020{\natexlab{c}})\citenamefont {Aaij} \emph
  {et~al.}}]{Aaij:2020gdg}%
  \BibitemOpen
  \bibfield  {author} {\bibinfo {author} {\bibfnamefont {R.}~\bibnamefont
  {Aaij}} \emph {et~al.} (\bibinfo {collaboration} {LHCb}),\ }\href@noop {} {\
  (\bibinfo {year} {2020}{\natexlab{c}})},\ \Eprint
  {http://arxiv.org/abs/2012.10380} {arXiv:2012.10380 [hep-ex]} \BibitemShut
  {NoStop}%
\bibitem [{\citenamefont {Wang}\ \emph
  {et~al.}(2020{\natexlab{a}})\citenamefont {Wang}, \citenamefont {Meng},\ and\
  \citenamefont {Zhu}}]{Wang:2019nvm}%
  \BibitemOpen
  \bibfield  {author} {\bibinfo {author} {\bibfnamefont {B.}~\bibnamefont
  {Wang}}, \bibinfo {author} {\bibfnamefont {L.}~\bibnamefont {Meng}}, \ and\
  \bibinfo {author} {\bibfnamefont {S.-L.}\ \bibnamefont {Zhu}},\ }\href
  {\doibase 10.1103/PhysRevD.101.034018} {\bibfield  {journal} {\bibinfo
  {journal} {Phys. Rev. D}\ }\textbf {\bibinfo {volume} {101}},\ \bibinfo
  {pages} {034018} (\bibinfo {year} {2020}{\natexlab{a}})},\ \Eprint
  {http://arxiv.org/abs/1912.12592} {arXiv:1912.12592 [hep-ph]} \BibitemShut
  {NoStop}%
\bibitem [{\citenamefont {Ablikim}\ \emph {et~al.}(2020)\citenamefont {Ablikim}
  \emph {et~al.}}]{Ablikim:2020hsk}%
  \BibitemOpen
  \bibfield  {author} {\bibinfo {author} {\bibfnamefont {M.}~\bibnamefont
  {Ablikim}} \emph {et~al.} (\bibinfo {collaboration} {BESIII}),\ }\href@noop
  {} {\  (\bibinfo {year} {2020})},\ \Eprint {http://arxiv.org/abs/2011.07855}
  {arXiv:2011.07855 [hep-ex]} \BibitemShut {NoStop}%
\bibitem [{\citenamefont {Meng}\ \emph {et~al.}(2020)\citenamefont {Meng},
  \citenamefont {Wang},\ and\ \citenamefont {Zhu}}]{Meng:2020ihj}%
  \BibitemOpen
  \bibfield  {author} {\bibinfo {author} {\bibfnamefont {L.}~\bibnamefont
  {Meng}}, \bibinfo {author} {\bibfnamefont {B.}~\bibnamefont {Wang}}, \ and\
  \bibinfo {author} {\bibfnamefont {S.-L.}\ \bibnamefont {Zhu}},\ }\href
  {\doibase 10.1103/PhysRevD.102.111502} {\bibfield  {journal} {\bibinfo
  {journal} {Phys. Rev. D}\ }\textbf {\bibinfo {volume} {102}},\ \bibinfo
  {pages} {111502} (\bibinfo {year} {2020})},\ \Eprint
  {http://arxiv.org/abs/2011.08656} {arXiv:2011.08656 [hep-ph]} \BibitemShut
  {NoStop}%
\bibitem [{\citenamefont {Wang}\ \emph
  {et~al.}(2021{\natexlab{a}})\citenamefont {Wang}, \citenamefont {Meng},\ and\
  \citenamefont {Zhu}}]{Wang:2020htx}%
  \BibitemOpen
  \bibfield  {author} {\bibinfo {author} {\bibfnamefont {B.}~\bibnamefont
  {Wang}}, \bibinfo {author} {\bibfnamefont {L.}~\bibnamefont {Meng}}, \ and\
  \bibinfo {author} {\bibfnamefont {S.-L.}\ \bibnamefont {Zhu}},\ }\href
  {\doibase 10.1103/PhysRevD.103.L021501} {\bibfield  {journal} {\bibinfo
  {journal} {Phys. Rev. D}\ }\textbf {\bibinfo {volume} {103}},\ \bibinfo
  {pages} {L021501} (\bibinfo {year} {2021}{\natexlab{a}})},\ \Eprint
  {http://arxiv.org/abs/2011.10922} {arXiv:2011.10922 [hep-ph]} \BibitemShut
  {NoStop}%
\bibitem [{\citenamefont {Wang}\ \emph
  {et~al.}(2020{\natexlab{b}})\citenamefont {Wang}, \citenamefont {Chen},\ and\
  \citenamefont {Chen}}]{Wang:2020rcx}%
  \BibitemOpen
  \bibfield  {author} {\bibinfo {author} {\bibfnamefont {Q.-N.}\ \bibnamefont
  {Wang}}, \bibinfo {author} {\bibfnamefont {W.}~\bibnamefont {Chen}}, \ and\
  \bibinfo {author} {\bibfnamefont {H.-X.}\ \bibnamefont {Chen}},\ }\href@noop
  {} {\  (\bibinfo {year} {2020}{\natexlab{b}})},\ \Eprint
  {http://arxiv.org/abs/2011.10495} {arXiv:2011.10495 [hep-ph]} \BibitemShut
  {NoStop}%
\bibitem [{\citenamefont {Sun}\ and\ \citenamefont {Xiao}(2020)}]{Sun:2020hjw}%
  \BibitemOpen
  \bibfield  {author} {\bibinfo {author} {\bibfnamefont {Z.-F.}\ \bibnamefont
  {Sun}}\ and\ \bibinfo {author} {\bibfnamefont {C.-W.}\ \bibnamefont {Xiao}},\
  }\href@noop {} {\  (\bibinfo {year} {2020})},\ \Eprint
  {http://arxiv.org/abs/2011.09404} {arXiv:2011.09404 [hep-ph]} \BibitemShut
  {NoStop}%
\bibitem [{\citenamefont {Du}\ \emph {et~al.}(2020)\citenamefont {Du},
  \citenamefont {Wang},\ and\ \citenamefont {Zhao}}]{Du:2020vwb}%
  \BibitemOpen
  \bibfield  {author} {\bibinfo {author} {\bibfnamefont {M.-C.}\ \bibnamefont
  {Du}}, \bibinfo {author} {\bibfnamefont {Q.}~\bibnamefont {Wang}}, \ and\
  \bibinfo {author} {\bibfnamefont {Q.}~\bibnamefont {Zhao}},\ }\href@noop {}
  {\  (\bibinfo {year} {2020})},\ \Eprint {http://arxiv.org/abs/2011.09225}
  {arXiv:2011.09225 [hep-ph]} \BibitemShut {NoStop}%
\bibitem [{\citenamefont {Cao}\ \emph {et~al.}(2021)\citenamefont {Cao},
  \citenamefont {Dai},\ and\ \citenamefont {Yang}}]{Cao:2020cfx}%
  \BibitemOpen
  \bibfield  {author} {\bibinfo {author} {\bibfnamefont {X.}~\bibnamefont
  {Cao}}, \bibinfo {author} {\bibfnamefont {J.-P.}\ \bibnamefont {Dai}}, \ and\
  \bibinfo {author} {\bibfnamefont {Z.}~\bibnamefont {Yang}},\ }\href {\doibase
  10.1140/epjc/s10052-021-08858-7} {\bibfield  {journal} {\bibinfo  {journal}
  {Eur. Phys. J. C}\ }\textbf {\bibinfo {volume} {81}},\ \bibinfo {pages} {184}
  (\bibinfo {year} {2021})},\ \Eprint {http://arxiv.org/abs/2011.09244}
  {arXiv:2011.09244 [hep-ph]} \BibitemShut {NoStop}%
\bibitem [{\citenamefont {Chen}\ and\ \citenamefont
  {Huang}(2021)}]{Chen:2020yvq}%
  \BibitemOpen
  \bibfield  {author} {\bibinfo {author} {\bibfnamefont {R.}~\bibnamefont
  {Chen}}\ and\ \bibinfo {author} {\bibfnamefont {Q.}~\bibnamefont {Huang}},\
  }\href {\doibase 10.1103/PhysRevD.103.034008} {\bibfield  {journal} {\bibinfo
   {journal} {Phys. Rev. D}\ }\textbf {\bibinfo {volume} {103}},\ \bibinfo
  {pages} {034008} (\bibinfo {year} {2021})},\ \Eprint
  {http://arxiv.org/abs/2011.09156} {arXiv:2011.09156 [hep-ph]} \BibitemShut
  {NoStop}%
\bibitem [{\citenamefont {Yang}\ \emph {et~al.}(2020)\citenamefont {Yang},
  \citenamefont {Cao}, \citenamefont {Guo}, \citenamefont {Nieves},\ and\
  \citenamefont {Valderrama}}]{Yang:2020nrt}%
  \BibitemOpen
  \bibfield  {author} {\bibinfo {author} {\bibfnamefont {Z.}~\bibnamefont
  {Yang}}, \bibinfo {author} {\bibfnamefont {X.}~\bibnamefont {Cao}}, \bibinfo
  {author} {\bibfnamefont {F.-K.}\ \bibnamefont {Guo}}, \bibinfo {author}
  {\bibfnamefont {J.}~\bibnamefont {Nieves}}, \ and\ \bibinfo {author}
  {\bibfnamefont {M.~P.}\ \bibnamefont {Valderrama}},\ }\href@noop {} {\
  (\bibinfo {year} {2020})},\ \Eprint {http://arxiv.org/abs/2011.08725}
  {arXiv:2011.08725 [hep-ph]} \BibitemShut {NoStop}%
\bibitem [{\citenamefont {Wang}\ \emph
  {et~al.}(2021{\natexlab{b}})\citenamefont {Wang}, \citenamefont {Zhou},
  \citenamefont {Liu},\ and\ \citenamefont {Matsuki}}]{Wang:2020kej}%
  \BibitemOpen
  \bibfield  {author} {\bibinfo {author} {\bibfnamefont {J.-Z.}\ \bibnamefont
  {Wang}}, \bibinfo {author} {\bibfnamefont {Q.-S.}\ \bibnamefont {Zhou}},
  \bibinfo {author} {\bibfnamefont {X.}~\bibnamefont {Liu}}, \ and\ \bibinfo
  {author} {\bibfnamefont {T.}~\bibnamefont {Matsuki}},\ }\href {\doibase
  10.1140/epjc/s10052-021-08877-4} {\bibfield  {journal} {\bibinfo  {journal}
  {Eur. Phys. J. C}\ }\textbf {\bibinfo {volume} {81}},\ \bibinfo {pages} {51}
  (\bibinfo {year} {2021}{\natexlab{b}})},\ \Eprint
  {http://arxiv.org/abs/2011.08628} {arXiv:2011.08628 [hep-ph]} \BibitemShut
  {NoStop}%
\bibitem [{\citenamefont {Azizi}\ and\ \citenamefont
  {Er}(2021)}]{Azizi:2020zyq}%
  \BibitemOpen
  \bibfield  {author} {\bibinfo {author} {\bibfnamefont {K.}~\bibnamefont
  {Azizi}}\ and\ \bibinfo {author} {\bibfnamefont {N.}~\bibnamefont {Er}},\
  }\href {\doibase 10.1140/epjc/s10052-021-08859-6} {\bibfield  {journal}
  {\bibinfo  {journal} {Eur. Phys. J. C}\ }\textbf {\bibinfo {volume} {81}},\
  \bibinfo {pages} {61} (\bibinfo {year} {2021})},\ \Eprint
  {http://arxiv.org/abs/2011.11488} {arXiv:2011.11488 [hep-ph]} \BibitemShut
  {NoStop}%
\bibitem [{\citenamefont {Jin}\ \emph {et~al.}(2020)\citenamefont {Jin},
  \citenamefont {Liu}, \citenamefont {Xue}, \citenamefont {Huang},\ and\
  \citenamefont {Ping}}]{Jin:2020yjn}%
  \BibitemOpen
  \bibfield  {author} {\bibinfo {author} {\bibfnamefont {X.}~\bibnamefont
  {Jin}}, \bibinfo {author} {\bibfnamefont {X.}~\bibnamefont {Liu}}, \bibinfo
  {author} {\bibfnamefont {Y.}~\bibnamefont {Xue}}, \bibinfo {author}
  {\bibfnamefont {H.}~\bibnamefont {Huang}}, \ and\ \bibinfo {author}
  {\bibfnamefont {J.}~\bibnamefont {Ping}},\ }\href@noop {} {\  (\bibinfo
  {year} {2020})},\ \Eprint {http://arxiv.org/abs/2011.12230} {arXiv:2011.12230
  [hep-ph]} \BibitemShut {NoStop}%
\bibitem [{\citenamefont {Wan}\ and\ \citenamefont {Qiao}(2020)}]{Wan:2020oxt}%
  \BibitemOpen
  \bibfield  {author} {\bibinfo {author} {\bibfnamefont {B.-D.}\ \bibnamefont
  {Wan}}\ and\ \bibinfo {author} {\bibfnamefont {C.-F.}\ \bibnamefont {Qiao}},\
  }\href@noop {} {\  (\bibinfo {year} {2020})},\ \Eprint
  {http://arxiv.org/abs/2011.08747} {arXiv:2011.08747 [hep-ph]} \BibitemShut
  {NoStop}%
\bibitem [{\citenamefont {Liu}\ \emph {et~al.}(2021)\citenamefont {Liu},
  \citenamefont {Pan},\ and\ \citenamefont {Geng}}]{Liu:2020hcv}%
  \BibitemOpen
  \bibfield  {author} {\bibinfo {author} {\bibfnamefont {M.-Z.}\ \bibnamefont
  {Liu}}, \bibinfo {author} {\bibfnamefont {Y.-W.}\ \bibnamefont {Pan}}, \ and\
  \bibinfo {author} {\bibfnamefont {L.-S.}\ \bibnamefont {Geng}},\ }\href
  {\doibase 10.1103/PhysRevD.103.034003} {\bibfield  {journal} {\bibinfo
  {journal} {Phys. Rev. D}\ }\textbf {\bibinfo {volume} {103}},\ \bibinfo
  {pages} {034003} (\bibinfo {year} {2021})},\ \Eprint
  {http://arxiv.org/abs/2011.07935} {arXiv:2011.07935 [hep-ph]} \BibitemShut
  {NoStop}%
\bibitem [{\citenamefont {Chen}(2021)}]{Chen:2020kco}%
  \BibitemOpen
  \bibfield  {author} {\bibinfo {author} {\bibfnamefont {R.}~\bibnamefont
  {Chen}},\ }\href {\doibase 10.1103/PhysRevD.103.054007} {\bibfield  {journal}
  {\bibinfo  {journal} {Phys. Rev. D}\ }\textbf {\bibinfo {volume} {103}},\
  \bibinfo {pages} {054007} (\bibinfo {year} {2021})},\ \Eprint
  {http://arxiv.org/abs/2011.07214} {arXiv:2011.07214 [hep-ph]} \BibitemShut
  {NoStop}%
\bibitem [{\citenamefont {Peng}\ \emph {et~al.}(2020)\citenamefont {Peng},
  \citenamefont {Yan}, \citenamefont {S\'anchez~S\'anchez},\ and\ \citenamefont
  {Valderrama}}]{Peng:2020hql}%
  \BibitemOpen
  \bibfield  {author} {\bibinfo {author} {\bibfnamefont {F.-Z.}\ \bibnamefont
  {Peng}}, \bibinfo {author} {\bibfnamefont {M.-J.}\ \bibnamefont {Yan}},
  \bibinfo {author} {\bibfnamefont {M.}~\bibnamefont {S\'anchez~S\'anchez}}, \
  and\ \bibinfo {author} {\bibfnamefont {M.~P.}\ \bibnamefont {Valderrama}},\
  }\href@noop {} {\  (\bibinfo {year} {2020})},\ \Eprint
  {http://arxiv.org/abs/2011.01915} {arXiv:2011.01915 [hep-ph]} \BibitemShut
  {NoStop}%
\bibitem [{\citenamefont {Chen}\ \emph {et~al.}(2020)\citenamefont {Chen},
  \citenamefont {Chen}, \citenamefont {Liu},\ and\ \citenamefont
  {Liu}}]{Chen:2020uif}%
  \BibitemOpen
  \bibfield  {author} {\bibinfo {author} {\bibfnamefont {H.-X.}\ \bibnamefont
  {Chen}}, \bibinfo {author} {\bibfnamefont {W.}~\bibnamefont {Chen}}, \bibinfo
  {author} {\bibfnamefont {X.}~\bibnamefont {Liu}}, \ and\ \bibinfo {author}
  {\bibfnamefont {X.-H.}\ \bibnamefont {Liu}},\ }\href@noop {} {\  (\bibinfo
  {year} {2020})},\ \Eprint {http://arxiv.org/abs/2011.01079} {arXiv:2011.01079
  [hep-ph]} \BibitemShut {NoStop}%
\bibitem [{\citenamefont {Shen}\ \emph {et~al.}(2020)\citenamefont {Shen},
  \citenamefont {Jing}, \citenamefont {Guo},\ and\ \citenamefont
  {Wu}}]{Shen:2020gpw}%
  \BibitemOpen
  \bibfield  {author} {\bibinfo {author} {\bibfnamefont {C.-W.}\ \bibnamefont
  {Shen}}, \bibinfo {author} {\bibfnamefont {H.-J.}\ \bibnamefont {Jing}},
  \bibinfo {author} {\bibfnamefont {F.-K.}\ \bibnamefont {Guo}}, \ and\
  \bibinfo {author} {\bibfnamefont {J.-J.}\ \bibnamefont {Wu}},\ }\href
  {\doibase 10.3390/sym12101611} {\bibfield  {journal} {\bibinfo  {journal}
  {Symmetry}\ }\textbf {\bibinfo {volume} {12}},\ \bibinfo {pages} {1611}
  (\bibinfo {year} {2020})},\ \Eprint {http://arxiv.org/abs/2008.09082}
  {arXiv:2008.09082 [hep-ph]} \BibitemShut {NoStop}%
\bibitem [{\citenamefont {Stancu}(2020)}]{Stancu:2020paw}%
  \BibitemOpen
  \bibfield  {author} {\bibinfo {author} {\bibfnamefont {F.}~\bibnamefont
  {Stancu}},\ }\href {\doibase 10.1103/PhysRevD.101.094007} {\bibfield
  {journal} {\bibinfo  {journal} {Phys. Rev. D}\ }\textbf {\bibinfo {volume}
  {101}},\ \bibinfo {pages} {094007} (\bibinfo {year} {2020})},\ \Eprint
  {http://arxiv.org/abs/2004.06009} {arXiv:2004.06009 [hep-ph]} \BibitemShut
  {NoStop}%
\bibitem [{\citenamefont {Xu}\ \emph {et~al.}(2020)\citenamefont {Xu},
  \citenamefont {Liu}, \citenamefont {Cui},\ and\ \citenamefont
  {Huang}}]{Xu:2020evn}%
  \BibitemOpen
  \bibfield  {author} {\bibinfo {author} {\bibfnamefont {Y.-J.}\ \bibnamefont
  {Xu}}, \bibinfo {author} {\bibfnamefont {Y.-L.}\ \bibnamefont {Liu}},
  \bibinfo {author} {\bibfnamefont {C.-Y.}\ \bibnamefont {Cui}}, \ and\
  \bibinfo {author} {\bibfnamefont {M.-Q.}\ \bibnamefont {Huang}},\ }\href@noop
  {} {\  (\bibinfo {year} {2020})},\ \Eprint {http://arxiv.org/abs/2011.14313}
  {arXiv:2011.14313 [hep-ph]} \BibitemShut {NoStop}%
\bibitem [{\citenamefont {Ikeno}\ \emph {et~al.}(2020)\citenamefont {Ikeno},
  \citenamefont {Molina},\ and\ \citenamefont {Oset}}]{Ikeno:2020csu}%
  \BibitemOpen
  \bibfield  {author} {\bibinfo {author} {\bibfnamefont {N.}~\bibnamefont
  {Ikeno}}, \bibinfo {author} {\bibfnamefont {R.}~\bibnamefont {Molina}}, \
  and\ \bibinfo {author} {\bibfnamefont {E.}~\bibnamefont {Oset}},\ }\href
  {\doibase 10.1016/j.physletb.2021.136120} {\  (\bibinfo {year} {2020}),\
  10.1016/j.physletb.2021.136120},\ \Eprint {http://arxiv.org/abs/2011.13425}
  {arXiv:2011.13425 [hep-ph]} \BibitemShut {NoStop}%
\bibitem [{\citenamefont {Nieves}\ and\ \citenamefont
  {Valderrama}(2012)}]{Nieves:2012tt}%
  \BibitemOpen
  \bibfield  {author} {\bibinfo {author} {\bibfnamefont {J.}~\bibnamefont
  {Nieves}}\ and\ \bibinfo {author} {\bibfnamefont {M.~P.}\ \bibnamefont
  {Valderrama}},\ }\href {\doibase 10.1103/PhysRevD.86.056004} {\bibfield
  {journal} {\bibinfo  {journal} {Phys. Rev. D}\ }\textbf {\bibinfo {volume}
  {86}},\ \bibinfo {pages} {056004} (\bibinfo {year} {2012})},\ \Eprint
  {http://arxiv.org/abs/1204.2790} {arXiv:1204.2790 [hep-ph]} \BibitemShut
  {NoStop}%
\bibitem [{\citenamefont {Guo}\ \emph {et~al.}(2013)\citenamefont {Guo},
  \citenamefont {Hidalgo-Duque}, \citenamefont {Nieves},\ and\ \citenamefont
  {Valderrama}}]{Guo:2013sya}%
  \BibitemOpen
  \bibfield  {author} {\bibinfo {author} {\bibfnamefont {F.-K.}\ \bibnamefont
  {Guo}}, \bibinfo {author} {\bibfnamefont {C.}~\bibnamefont {Hidalgo-Duque}},
  \bibinfo {author} {\bibfnamefont {J.}~\bibnamefont {Nieves}}, \ and\ \bibinfo
  {author} {\bibfnamefont {M.~P.}\ \bibnamefont {Valderrama}},\ }\href
  {\doibase 10.1103/PhysRevD.88.054007} {\bibfield  {journal} {\bibinfo
  {journal} {Phys. Rev. D}\ }\textbf {\bibinfo {volume} {88}},\ \bibinfo
  {pages} {054007} (\bibinfo {year} {2013})},\ \Eprint
  {http://arxiv.org/abs/1303.6608} {arXiv:1303.6608 [hep-ph]} \BibitemShut
  {NoStop}%
\bibitem [{\citenamefont {Baru}\ \emph {et~al.}(2016)\citenamefont {Baru},
  \citenamefont {Epelbaum}, \citenamefont {Filin}, \citenamefont {Hanhart},
  \citenamefont {Mei\ss{}ner},\ and\ \citenamefont {Nefediev}}]{Baru:2016iwj}%
  \BibitemOpen
  \bibfield  {author} {\bibinfo {author} {\bibfnamefont {V.}~\bibnamefont
  {Baru}}, \bibinfo {author} {\bibfnamefont {E.}~\bibnamefont {Epelbaum}},
  \bibinfo {author} {\bibfnamefont {A.~A.}\ \bibnamefont {Filin}}, \bibinfo
  {author} {\bibfnamefont {C.}~\bibnamefont {Hanhart}}, \bibinfo {author}
  {\bibfnamefont {U.-G.}\ \bibnamefont {Mei\ss{}ner}}, \ and\ \bibinfo {author}
  {\bibfnamefont {A.~V.}\ \bibnamefont {Nefediev}},\ }\href {\doibase
  10.1016/j.physletb.2016.10.008} {\bibfield  {journal} {\bibinfo  {journal}
  {Phys. Lett. B}\ }\textbf {\bibinfo {volume} {763}},\ \bibinfo {pages} {20}
  (\bibinfo {year} {2016})},\ \Eprint {http://arxiv.org/abs/1605.09649}
  {arXiv:1605.09649 [hep-ph]} \BibitemShut {NoStop}%
\bibitem [{\citenamefont {Meng}\ \emph {et~al.}(2019)\citenamefont {Meng},
  \citenamefont {Wang}, \citenamefont {Wang},\ and\ \citenamefont
  {Zhu}}]{Meng:2019ilv}%
  \BibitemOpen
  \bibfield  {author} {\bibinfo {author} {\bibfnamefont {L.}~\bibnamefont
  {Meng}}, \bibinfo {author} {\bibfnamefont {B.}~\bibnamefont {Wang}}, \bibinfo
  {author} {\bibfnamefont {G.-J.}\ \bibnamefont {Wang}}, \ and\ \bibinfo
  {author} {\bibfnamefont {S.-L.}\ \bibnamefont {Zhu}},\ }\href {\doibase
  10.1103/PhysRevD.100.014031} {\bibfield  {journal} {\bibinfo  {journal}
  {Phys. Rev. D}\ }\textbf {\bibinfo {volume} {100}},\ \bibinfo {pages}
  {014031} (\bibinfo {year} {2019})},\ \Eprint
  {http://arxiv.org/abs/1905.04113} {arXiv:1905.04113 [hep-ph]} \BibitemShut
  {NoStop}%
\bibitem [{\citenamefont {Wang}\ \emph {et~al.}(2019)\citenamefont {Wang},
  \citenamefont {Meng},\ and\ \citenamefont {Zhu}}]{Wang:2019ato}%
  \BibitemOpen
  \bibfield  {author} {\bibinfo {author} {\bibfnamefont {B.}~\bibnamefont
  {Wang}}, \bibinfo {author} {\bibfnamefont {L.}~\bibnamefont {Meng}}, \ and\
  \bibinfo {author} {\bibfnamefont {S.-L.}\ \bibnamefont {Zhu}},\ }\href
  {\doibase 10.1007/JHEP11(2019)108} {\bibfield  {journal} {\bibinfo  {journal}
  {JHEP}\ }\textbf {\bibinfo {volume} {11}},\ \bibinfo {pages} {108} (\bibinfo
  {year} {2019})},\ \Eprint {http://arxiv.org/abs/1909.13054} {arXiv:1909.13054
  [hep-ph]} \BibitemShut {NoStop}%
\bibitem [{\citenamefont {Zyla}\ \emph {et~al.}(2020)\citenamefont {Zyla} \emph
  {et~al.}}]{Zyla:2020zbs}%
  \BibitemOpen
  \bibfield  {author} {\bibinfo {author} {\bibfnamefont {P.~A.}\ \bibnamefont
  {Zyla}} \emph {et~al.} (\bibinfo {collaboration} {Particle Data Group}),\
  }\href {\doibase 10.1093/ptep/ptaa104} {\bibfield  {journal} {\bibinfo
  {journal} {PTEP}\ }\textbf {\bibinfo {volume} {2020}},\ \bibinfo {pages}
  {083C01} (\bibinfo {year} {2020})}\BibitemShut {NoStop}%
\bibitem [{\citenamefont {del Amo~Sanchez}\ \emph {et~al.}(2010)\citenamefont
  {del Amo~Sanchez} \emph {et~al.}}]{delAmoSanchez:2010jr}%
  \BibitemOpen
  \bibfield  {author} {\bibinfo {author} {\bibfnamefont {P.}~\bibnamefont {del
  Amo~Sanchez}} \emph {et~al.} (\bibinfo {collaboration} {BaBar}),\ }\href
  {\doibase 10.1103/PhysRevD.82.011101} {\bibfield  {journal} {\bibinfo
  {journal} {Phys. Rev. D}\ }\textbf {\bibinfo {volume} {82}},\ \bibinfo
  {pages} {011101} (\bibinfo {year} {2010})},\ \Eprint
  {http://arxiv.org/abs/1005.5190} {arXiv:1005.5190 [hep-ex]} \BibitemShut
  {NoStop}%
\bibitem [{\citenamefont {Li}\ and\ \citenamefont {Zhu}(2012)}]{Li:2012cs}%
  \BibitemOpen
  \bibfield  {author} {\bibinfo {author} {\bibfnamefont {N.}~\bibnamefont
  {Li}}\ and\ \bibinfo {author} {\bibfnamefont {S.-L.}\ \bibnamefont {Zhu}},\
  }\href {\doibase 10.1103/PhysRevD.86.074022} {\bibfield  {journal} {\bibinfo
  {journal} {Phys. Rev. D}\ }\textbf {\bibinfo {volume} {86}},\ \bibinfo
  {pages} {074022} (\bibinfo {year} {2012})},\ \Eprint
  {http://arxiv.org/abs/1207.3954} {arXiv:1207.3954 [hep-ph]} \BibitemShut
  {NoStop}%
\bibitem [{\citenamefont {Fleming}\ \emph {et~al.}(2007)\citenamefont
  {Fleming}, \citenamefont {Kusunoki}, \citenamefont {Mehen},\ and\
  \citenamefont {van Kolck}}]{Fleming:2007rp}%
  \BibitemOpen
  \bibfield  {author} {\bibinfo {author} {\bibfnamefont {S.}~\bibnamefont
  {Fleming}}, \bibinfo {author} {\bibfnamefont {M.}~\bibnamefont {Kusunoki}},
  \bibinfo {author} {\bibfnamefont {T.}~\bibnamefont {Mehen}}, \ and\ \bibinfo
  {author} {\bibfnamefont {U.}~\bibnamefont {van Kolck}},\ }\href {\doibase
  10.1103/PhysRevD.76.034006} {\bibfield  {journal} {\bibinfo  {journal} {Phys.
  Rev. D}\ }\textbf {\bibinfo {volume} {76}},\ \bibinfo {pages} {034006}
  (\bibinfo {year} {2007})},\ \Eprint {http://arxiv.org/abs/hep-ph/0703168}
  {arXiv:hep-ph/0703168} \BibitemShut {NoStop}%
\bibitem [{\citenamefont {Hidalgo-Duque}\ \emph {et~al.}(2013)\citenamefont
  {Hidalgo-Duque}, \citenamefont {Nieves},\ and\ \citenamefont
  {Valderrama}}]{HidalgoDuque:2012pq}%
  \BibitemOpen
  \bibfield  {author} {\bibinfo {author} {\bibfnamefont {C.}~\bibnamefont
  {Hidalgo-Duque}}, \bibinfo {author} {\bibfnamefont {J.}~\bibnamefont
  {Nieves}}, \ and\ \bibinfo {author} {\bibfnamefont {M.~P.}\ \bibnamefont
  {Valderrama}},\ }\href {\doibase 10.1103/PhysRevD.87.076006} {\bibfield
  {journal} {\bibinfo  {journal} {Phys. Rev. D}\ }\textbf {\bibinfo {volume}
  {87}},\ \bibinfo {pages} {076006} (\bibinfo {year} {2013})},\ \Eprint
  {http://arxiv.org/abs/1210.5431} {arXiv:1210.5431 [hep-ph]} \BibitemShut
  {NoStop}%
\bibitem [{\citenamefont {Gamermann}\ and\ \citenamefont
  {Oset}(2009)}]{Gamermann:2009fv}%
  \BibitemOpen
  \bibfield  {author} {\bibinfo {author} {\bibfnamefont {D.}~\bibnamefont
  {Gamermann}}\ and\ \bibinfo {author} {\bibfnamefont {E.}~\bibnamefont
  {Oset}},\ }\href {\doibase 10.1103/PhysRevD.80.014003} {\bibfield  {journal}
  {\bibinfo  {journal} {Phys. Rev. D}\ }\textbf {\bibinfo {volume} {80}},\
  \bibinfo {pages} {014003} (\bibinfo {year} {2009})},\ \Eprint
  {http://arxiv.org/abs/0905.0402} {arXiv:0905.0402 [hep-ph]} \BibitemShut
  {NoStop}%
\bibitem [{\citenamefont {Gamermann}\ \emph {et~al.}(2010)\citenamefont
  {Gamermann}, \citenamefont {Nieves}, \citenamefont {Oset},\ and\
  \citenamefont {Ruiz~Arriola}}]{Gamermann:2009uq}%
  \BibitemOpen
  \bibfield  {author} {\bibinfo {author} {\bibfnamefont {D.}~\bibnamefont
  {Gamermann}}, \bibinfo {author} {\bibfnamefont {J.}~\bibnamefont {Nieves}},
  \bibinfo {author} {\bibfnamefont {E.}~\bibnamefont {Oset}}, \ and\ \bibinfo
  {author} {\bibfnamefont {E.}~\bibnamefont {Ruiz~Arriola}},\ }\href {\doibase
  10.1103/PhysRevD.81.014029} {\bibfield  {journal} {\bibinfo  {journal} {Phys.
  Rev. D}\ }\textbf {\bibinfo {volume} {81}},\ \bibinfo {pages} {014029}
  (\bibinfo {year} {2010})},\ \Eprint {http://arxiv.org/abs/0911.4407}
  {arXiv:0911.4407 [hep-ph]} \BibitemShut {NoStop}%
\bibitem [{\citenamefont {Cohen}\ \emph {et~al.}(2004)\citenamefont {Cohen},
  \citenamefont {Gelman},\ and\ \citenamefont {van Kolck}}]{Cohen:2004kf}%
  \BibitemOpen
  \bibfield  {author} {\bibinfo {author} {\bibfnamefont {T.~D.}\ \bibnamefont
  {Cohen}}, \bibinfo {author} {\bibfnamefont {B.~A.}\ \bibnamefont {Gelman}}, \
  and\ \bibinfo {author} {\bibfnamefont {U.}~\bibnamefont {van Kolck}},\ }\href
  {\doibase 10.1016/j.physletb.2004.03.020} {\bibfield  {journal} {\bibinfo
  {journal} {Phys. Lett. B}\ }\textbf {\bibinfo {volume} {588}},\ \bibinfo
  {pages} {57} (\bibinfo {year} {2004})},\ \Eprint
  {http://arxiv.org/abs/nucl-th/0402054} {arXiv:nucl-th/0402054} \BibitemShut
  {NoStop}%
\bibitem [{\citenamefont {Braaten}\ and\ \citenamefont
  {Kusunoki}(2005)}]{Braaten:2005ai}%
  \BibitemOpen
  \bibfield  {author} {\bibinfo {author} {\bibfnamefont {E.}~\bibnamefont
  {Braaten}}\ and\ \bibinfo {author} {\bibfnamefont {M.}~\bibnamefont
  {Kusunoki}},\ }\href {\doibase 10.1103/PhysRevD.72.054022} {\bibfield
  {journal} {\bibinfo  {journal} {Phys. Rev. D}\ }\textbf {\bibinfo {volume}
  {72}},\ \bibinfo {pages} {054022} (\bibinfo {year} {2005})},\ \Eprint
  {http://arxiv.org/abs/hep-ph/0507163} {arXiv:hep-ph/0507163} \BibitemShut
  {NoStop}%
\bibitem [{\citenamefont {Dong}\ \emph {et~al.}(2020)\citenamefont {Dong},
  \citenamefont {Guo},\ and\ \citenamefont {Zou}}]{Dong:2020hxe}%
  \BibitemOpen
  \bibfield  {author} {\bibinfo {author} {\bibfnamefont {X.-K.}\ \bibnamefont
  {Dong}}, \bibinfo {author} {\bibfnamefont {F.-K.}\ \bibnamefont {Guo}}, \
  and\ \bibinfo {author} {\bibfnamefont {B.-S.}\ \bibnamefont {Zou}},\
  }\href@noop {} {\  (\bibinfo {year} {2020})},\ \Eprint
  {http://arxiv.org/abs/2011.14517} {arXiv:2011.14517 [hep-ph]} \BibitemShut
  {NoStop}%
\bibitem [{\citenamefont {Ablikim}\ \emph
  {et~al.}(2015{\natexlab{a}})\citenamefont {Ablikim} \emph
  {et~al.}}]{Ablikim:2015gda}%
  \BibitemOpen
  \bibfield  {author} {\bibinfo {author} {\bibfnamefont {M.}~\bibnamefont
  {Ablikim}} \emph {et~al.} (\bibinfo {collaboration} {BESIII}),\ }\href
  {\doibase 10.1103/PhysRevLett.115.222002} {\bibfield  {journal} {\bibinfo
  {journal} {Phys. Rev. Lett.}\ }\textbf {\bibinfo {volume} {115}},\ \bibinfo
  {pages} {222002} (\bibinfo {year} {2015}{\natexlab{a}})},\ \Eprint
  {http://arxiv.org/abs/1509.05620} {arXiv:1509.05620 [hep-ex]} \BibitemShut
  {NoStop}%
\bibitem [{\citenamefont {Ablikim}\ \emph
  {et~al.}(2015{\natexlab{b}})\citenamefont {Ablikim} \emph
  {et~al.}}]{Ablikim:2015tbp}%
  \BibitemOpen
  \bibfield  {author} {\bibinfo {author} {\bibfnamefont {M.}~\bibnamefont
  {Ablikim}} \emph {et~al.} (\bibinfo {collaboration} {BESIII}),\ }\href
  {\doibase 10.1103/PhysRevLett.115.112003} {\bibfield  {journal} {\bibinfo
  {journal} {Phys. Rev. Lett.}\ }\textbf {\bibinfo {volume} {115}},\ \bibinfo
  {pages} {112003} (\bibinfo {year} {2015}{\natexlab{b}})},\ \Eprint
  {http://arxiv.org/abs/1506.06018} {arXiv:1506.06018 [hep-ex]} \BibitemShut
  {NoStop}%
\bibitem [{\citenamefont {Ablikim}\ \emph
  {et~al.}(2015{\natexlab{c}})\citenamefont {Ablikim} \emph
  {et~al.}}]{Ablikim:2015vvn}%
  \BibitemOpen
  \bibfield  {author} {\bibinfo {author} {\bibfnamefont {M.}~\bibnamefont
  {Ablikim}} \emph {et~al.} (\bibinfo {collaboration} {BESIII}),\ }\href
  {\doibase 10.1103/PhysRevLett.115.182002} {\bibfield  {journal} {\bibinfo
  {journal} {Phys. Rev. Lett.}\ }\textbf {\bibinfo {volume} {115}},\ \bibinfo
  {pages} {182002} (\bibinfo {year} {2015}{\natexlab{c}})},\ \Eprint
  {http://arxiv.org/abs/1507.02404} {arXiv:1507.02404 [hep-ex]} \BibitemShut
  {NoStop}%
\bibitem [{\citenamefont {Braaten}\ and\ \citenamefont
  {Kusunoki}(2004)}]{Braaten:2003he}%
  \BibitemOpen
  \bibfield  {author} {\bibinfo {author} {\bibfnamefont {E.}~\bibnamefont
  {Braaten}}\ and\ \bibinfo {author} {\bibfnamefont {M.}~\bibnamefont
  {Kusunoki}},\ }\href {\doibase 10.1103/PhysRevD.69.074005} {\bibfield
  {journal} {\bibinfo  {journal} {Phys. Rev. D}\ }\textbf {\bibinfo {volume}
  {69}},\ \bibinfo {pages} {074005} (\bibinfo {year} {2004})},\ \Eprint
  {http://arxiv.org/abs/hep-ph/0311147} {arXiv:hep-ph/0311147} \BibitemShut
  {NoStop}%
\bibitem [{\citenamefont {Molina}\ and\ \citenamefont
  {Oset}(2009)}]{Molina:2009ct}%
  \BibitemOpen
  \bibfield  {author} {\bibinfo {author} {\bibfnamefont {R.}~\bibnamefont
  {Molina}}\ and\ \bibinfo {author} {\bibfnamefont {E.}~\bibnamefont {Oset}},\
  }\href {\doibase 10.1103/PhysRevD.80.114013} {\bibfield  {journal} {\bibinfo
  {journal} {Phys. Rev. D}\ }\textbf {\bibinfo {volume} {80}},\ \bibinfo
  {pages} {114013} (\bibinfo {year} {2009})},\ \Eprint
  {http://arxiv.org/abs/0907.3043} {arXiv:0907.3043 [hep-ph]} \BibitemShut
  {NoStop}%
\bibitem [{\citenamefont {Lees}\ \emph {et~al.}(2012)\citenamefont {Lees} \emph
  {et~al.}}]{Lees:2012xs}%
  \BibitemOpen
  \bibfield  {author} {\bibinfo {author} {\bibfnamefont {J.~P.}\ \bibnamefont
  {Lees}} \emph {et~al.} (\bibinfo {collaboration} {BaBar}),\ }\href {\doibase
  10.1103/PhysRevD.86.072002} {\bibfield  {journal} {\bibinfo  {journal} {Phys.
  Rev. D}\ }\textbf {\bibinfo {volume} {86}},\ \bibinfo {pages} {072002}
  (\bibinfo {year} {2012})},\ \Eprint {http://arxiv.org/abs/1207.2651}
  {arXiv:1207.2651 [hep-ex]} \BibitemShut {NoStop}%
\bibitem [{\citenamefont {Zhou}\ \emph {et~al.}(2015)\citenamefont {Zhou},
  \citenamefont {Xiao},\ and\ \citenamefont {Zhou}}]{Zhou:2015uva}%
  \BibitemOpen
  \bibfield  {author} {\bibinfo {author} {\bibfnamefont {Z.-Y.}\ \bibnamefont
  {Zhou}}, \bibinfo {author} {\bibfnamefont {Z.}~\bibnamefont {Xiao}}, \ and\
  \bibinfo {author} {\bibfnamefont {H.-Q.}\ \bibnamefont {Zhou}},\ }\href
  {\doibase 10.1103/PhysRevLett.115.022001} {\bibfield  {journal} {\bibinfo
  {journal} {Phys. Rev. Lett.}\ }\textbf {\bibinfo {volume} {115}},\ \bibinfo
  {pages} {022001} (\bibinfo {year} {2015})},\ \Eprint
  {http://arxiv.org/abs/1501.00879} {arXiv:1501.00879 [hep-ph]} \BibitemShut
  {NoStop}%
\bibitem [{\citenamefont {Li}\ and\ \citenamefont
  {Voloshin}(2015)}]{Li:2015iga}%
  \BibitemOpen
  \bibfield  {author} {\bibinfo {author} {\bibfnamefont {X.}~\bibnamefont
  {Li}}\ and\ \bibinfo {author} {\bibfnamefont {M.~B.}\ \bibnamefont
  {Voloshin}},\ }\href {\doibase 10.1103/PhysRevD.91.114014} {\bibfield
  {journal} {\bibinfo  {journal} {Phys. Rev. D}\ }\textbf {\bibinfo {volume}
  {91}},\ \bibinfo {pages} {114014} (\bibinfo {year} {2015})},\ \Eprint
  {http://arxiv.org/abs/1503.04431} {arXiv:1503.04431 [hep-ph]} \BibitemShut
  {NoStop}%
\bibitem [{\citenamefont {Lebed}\ and\ \citenamefont
  {Polosa}(2016)}]{Lebed:2016yvr}%
  \BibitemOpen
  \bibfield  {author} {\bibinfo {author} {\bibfnamefont {R.~F.}\ \bibnamefont
  {Lebed}}\ and\ \bibinfo {author} {\bibfnamefont {A.~D.}\ \bibnamefont
  {Polosa}},\ }\href {\doibase 10.1103/PhysRevD.93.094024} {\bibfield
  {journal} {\bibinfo  {journal} {Phys. Rev. D}\ }\textbf {\bibinfo {volume}
  {93}},\ \bibinfo {pages} {094024} (\bibinfo {year} {2016})},\ \Eprint
  {http://arxiv.org/abs/1602.08421} {arXiv:1602.08421 [hep-ph]} \BibitemShut
  {NoStop}%
\bibitem [{\citenamefont {Liu}\ \emph {et~al.}(2010)\citenamefont {Liu},
  \citenamefont {Luo},\ and\ \citenamefont {Sun}}]{Liu:2009fe}%
  \BibitemOpen
  \bibfield  {author} {\bibinfo {author} {\bibfnamefont {X.}~\bibnamefont
  {Liu}}, \bibinfo {author} {\bibfnamefont {Z.-G.}\ \bibnamefont {Luo}}, \ and\
  \bibinfo {author} {\bibfnamefont {Z.-F.}\ \bibnamefont {Sun}},\ }\href
  {\doibase 10.1103/PhysRevLett.104.122001} {\bibfield  {journal} {\bibinfo
  {journal} {Phys. Rev. Lett.}\ }\textbf {\bibinfo {volume} {104}},\ \bibinfo
  {pages} {122001} (\bibinfo {year} {2010})},\ \Eprint
  {http://arxiv.org/abs/0911.3694} {arXiv:0911.3694 [hep-ph]} \BibitemShut
  {NoStop}%
\bibitem [{\citenamefont {Gamermann}\ \emph {et~al.}(2007)\citenamefont
  {Gamermann}, \citenamefont {Oset}, \citenamefont {Strottman},\ and\
  \citenamefont {Vicente~Vacas}}]{Gamermann:2006nm}%
  \BibitemOpen
  \bibfield  {author} {\bibinfo {author} {\bibfnamefont {D.}~\bibnamefont
  {Gamermann}}, \bibinfo {author} {\bibfnamefont {E.}~\bibnamefont {Oset}},
  \bibinfo {author} {\bibfnamefont {D.}~\bibnamefont {Strottman}}, \ and\
  \bibinfo {author} {\bibfnamefont {M.~J.}\ \bibnamefont {Vicente~Vacas}},\
  }\href {\doibase 10.1103/PhysRevD.76.074016} {\bibfield  {journal} {\bibinfo
  {journal} {Phys. Rev. D}\ }\textbf {\bibinfo {volume} {76}},\ \bibinfo
  {pages} {074016} (\bibinfo {year} {2007})},\ \Eprint
  {http://arxiv.org/abs/hep-ph/0612179} {arXiv:hep-ph/0612179} \BibitemShut
  {NoStop}%
\bibitem [{\citenamefont {Aaij}\ \emph
  {et~al.}(2017{\natexlab{a}})\citenamefont {Aaij} \emph
  {et~al.}}]{Aaij:2016iza}%
  \BibitemOpen
  \bibfield  {author} {\bibinfo {author} {\bibfnamefont {R.}~\bibnamefont
  {Aaij}} \emph {et~al.} (\bibinfo {collaboration} {LHCb}),\ }\href {\doibase
  10.1103/PhysRevLett.118.022003} {\bibfield  {journal} {\bibinfo  {journal}
  {Phys. Rev. Lett.}\ }\textbf {\bibinfo {volume} {118}},\ \bibinfo {pages}
  {022003} (\bibinfo {year} {2017}{\natexlab{a}})},\ \Eprint
  {http://arxiv.org/abs/1606.07895} {arXiv:1606.07895 [hep-ex]} \BibitemShut
  {NoStop}%
\bibitem [{\citenamefont {Aaij}\ \emph
  {et~al.}(2017{\natexlab{b}})\citenamefont {Aaij} \emph
  {et~al.}}]{Aaij:2016nsc}%
  \BibitemOpen
  \bibfield  {author} {\bibinfo {author} {\bibfnamefont {R.}~\bibnamefont
  {Aaij}} \emph {et~al.} (\bibinfo {collaboration} {LHCb}),\ }\href {\doibase
  10.1103/PhysRevD.95.012002} {\bibfield  {journal} {\bibinfo  {journal} {Phys.
  Rev. D}\ }\textbf {\bibinfo {volume} {95}},\ \bibinfo {pages} {012002}
  (\bibinfo {year} {2017}{\natexlab{b}})},\ \Eprint
  {http://arxiv.org/abs/1606.07898} {arXiv:1606.07898 [hep-ex]} \BibitemShut
  {NoStop}%
\end{thebibliography}%

\end{document}